\documentclass[useAMS,usenatbib]{mn2e}
\pdfoutput=1
\usepackage{graphicx}
\usepackage{color}
\usepackage{psfig}
\usepackage{fancyhdr,rotating,aas_macros,amssymb,amsmath}
\usepackage{mathptmx}
\usepackage[T1]{fontenc}
\newcommand{\lya}{\mbox{$\rmn{Ly}\alpha$}}

\newcommand{\ha}{\mbox{$\rmn{H}\alpha$}}
\newcommand{\hb}{\mbox{$\rmn{H}\beta$}}
\newcommand{\galform}{\texttt{GALFORM}}

\newcommand{\fesc}{\mbox{$\rm{f}_{esc}$}}

\newcommand{\lunits}{\mbox{$[\rmn{erg} \ \rmn{s}^{-1} \ \rmn{h}^{-2}]$}}		%Luminosity units
\newcommand{\lha}{\mbox{$L_{\rmn{H}\alpha}$}}						%Lum. abbreviation
\newcommand{\sag}{\texttt{SAG}}
\newcommand{\mapp}{\texttt{MAPPINGS-III}}

\newcommand{\oii}{\mbox{$\rm [O II] \lambda 3727$}}
\newcommand{\oiii}{\mbox{$\rm [O III] \lambda 5007$}}
\newcommand{\cii}{\mbox{$\rm [C II] \lambda 158 \mu m$}}
\newcommand{\nii}{\mbox{$\rm [N II] \lambda 205 \mu m$}}
\newcommand{\niiott}{\mbox{$\rm [N II] \lambda 122 \mu m$}}
\newcommand{\oiiiee}{\mbox{$\rm [O III] \lambda 88 \mu m$}}

\newcommand{\aOII}{298.57}
\newcommand{\bOII}{-26.10}
\newcommand{\cOII}{1.33}
\newcommand{\dOII}{-0.01}
\newcommand{\eOII}{-15.50}
\newcommand{\fOII}{0.20}

\newcommand{\aOIII}{473.99}
\newcommand{\bOIII}{-190.35}
\newcommand{\cOIII}{9.29}
\newcommand{\dOIII}{-0.11}
\newcommand{\eOIII}{-24.46}
\newcommand{\fOIII}{0.31}

\newcommand{\aNII}{17.73}
\newcommand{\bNII}{46.98}
\newcommand{\cNII}{-2.21}
\newcommand{\dNII}{0.02}
\newcommand{\eNII}{-1.87}
\newcommand{\fNII}{0.03}

\newcommand{\aHa}{-41.48}

\title[The nebular emission of star-forming galaxies]
{The nebular emission of star-forming galaxies in a hierarchical universe} 
\author [A. Orsi et al.]
{ 
\'Alvaro Orsi$^{1,2}$\thanks{Email: aaorsi@astro.puc.cl},  
Nelson Padilla $^{1,2}$,
Brent Groves $^3$,  Sof\'ia Cora $^{4,5,6}$,
Tom\'as Tecce $^{1,2}$,\newauthor
Ignacio Gargiulo $^{4}$ and 
Andr\'es Ruiz $^{6,7,8}$
\vspace{0.01cm}\\ 
1. Instituto de Astrof\'isica, Pontificia Universidad Cat\'olica, Av. Vicu\~na Mackenna 4860, Santiago, Chile. \\
2. Centro de Astro-Ingenier\'ia, Pontificia Universidad Cat\'olica, Av. Vicu\~na Mackenna 4860, Santiago, Chile.\\
3. Max Planck Institute for Astronomy, K\"{o}nigstuhl 17 D-69117 Heidelberg, Germany \\
4. Instituto de Astrof\'isica de La Plata (CCT La Plata, CONICET, UNLP), Paseo del Bosque s/n, B1900FWA, La Plata, Argentina.\\
5. Facultad de Ciencias Astron\'omicas y Geof\'{\i}sicas, Universidad Nacional de La Plata, Paseo del Bosque s/n, B1900FWA, La Plata, Argentina.\\
6. Consejo Nacional de Investigaciones Cient\'{\i}ficas y T\'ecnicas, Rivadavia 1917, C1033AAJ Buenos Aires, Argentina.\\
7. Instituto de Astronom\'{\i}a Te\'orica y Experimental (CCT C\'ordoba, CONICET, UNC), Laprida 854, X5000BGR, C\'ordoba, Argentina.\\
8. Observatorio Astron\'omico de C\'ordoba, Universidad Nacional de C\'ordoba, Laprida 854, X5000BGR, C\'ordoba, Argentina.
\\
}
\begin{document}
\maketitle
\begin{abstract}
{ Galaxy surveys targeting emission lines are characterising the evolution of star-forming galaxies, 
but there is still little theoretical progress in modelling their physical properties.
We predict nebular emission from star-forming galaxies within a cosmological galaxy formation model.
Emission lines are computed by combining the semi-analytical model \sag\ with the photoionisation code \mapp. 
We characterise the interstellar medium (ISM) of galaxies 
by relating the ionisation parameter of gas in galaxies to their cold gas metallicity, obtaining a reasonable agreement with the 
observed \ha, \oii, \oiii\ luminosity functions, and the the BPT diagram for local star-forming galaxies.
The average ionisation parameter is found to increase towards low star-formation rates and high redshifts, 
consistent with recent observational results.
The predicted link between different emission lines and their associated star-formation rates is studied by presenting scaling relations to relate them.
Our model predicts that emission line galaxies have modest clustering bias, and thus reside in dark matter haloes of masses below
$M_{\rm halo} \lesssim 10^{12} {[\rm h^{-1} M_{\odot}]}$. Finally,
we exploit our modelling technique to predict galaxy number counts up to $z\sim 10$ 
by targeting far-infrared (FIR) emission lines detectable with submillimetre facilities.}
\end{abstract}


\section{Introduction}

Emission lines are a common feature in the spectra of galaxies. A number
of physical processes occurring in the interstellar medium (ISM) of galaxies 
can be responsible for their production, such as the recombination of ionised gas, 
collisional excitation or fluorescent excitation \citep{osterbrock89,stasinska07}. 
Their detection allows the exploration of the physical properties of the medium
from which they are produced. Active galactic nuclei (AGN), for instance, 
can photoionise and shock-heat the gas in the narrow-line region surrounding
the accretion disks of supermassive black holes \citep{groves04a,groves04b}. 
Also, cold gas falling into the galaxy's potential well might radiate its energy 
away via collisional excitation \citep{dijkstra09}.

Likewise, star-forming activity in galaxies can be traced by the 
detection of emission lines \citep{kennicutt83,calzetti12a}. 
Young, massive stars produce photoionising radiation
which is absorbed by the neutral gas of the ISM leading to the production of emission lines.
Given the short life-times of massive stars
(of the order of a few Myr), emission lines indicate the occurrence of recent episodes 
of star formation. On the contrary, continuum-based tracers of star forming (SF) activity, 
such as the UV continuum, are tracing the star formation rate (SFR) over a much larger time-scale of the order of $\sim$100 Myr.
Regardless of the technique used, mapping the cosmic evolution of star formation to understand the formation and evolution of galaxies 
has become an important challenge in extragalactic astrophysics { \citep{madau98,hopkins04,madau14}}.  

The link between emission line fluxes and the properties of the gas in the ISM have naturally made their 
detection a common tool in extragalactic astrophysics.  
Theoretical modelling of emission lines has focused mostly in providing a tool 
to learn about the dynamics and structure of the gas in galaxies. 
Typically, the interpretation of line fluxes and line ratios is done by making use of a photoionisation radiative transfer code 
\citep{ferland98,dopita00,kewley01}. Line ratios and fluxes can then 
be translated into gas metallicities, temperatures, densities and the ionising structure of the gas 
\citep[e.g.][]{kewley08,levesque10}. By using a number of ad-hoc assumptions, it is also possible 
to include nebular emission into a stellar population synthesis code \citep{charlot01,panuzzo03}.
Individual objects can be studied in great detail when both optical and infrared emission lines are available, constraining the 
properties of the multi-phase ISM and the contribution of each ISM component to the total emission 
line budget \citep[e.g.][]{abel05,cormier12}.  

Analysis of the emission line budget has also been done in large galaxy datasets. In the local spectroscopic sample of 
galaxies from the Sloan Digital Sky Survey (SDSS), \citet{tremonti04} made use
of several line ratios finding a global relation between the gas metallicity and the stellar mass of galaxies. 
The exact shape and normalisation of this relation is somewhat controversial, since metallicity diagnostics
using different line ratios can lead to different values of metallicity \citep{kewley08,cullen13}. Despite the 
uncertainty in the exact value of the gas metallicity inferred by measuring line ratios, 
it is clear that the mass-metallicity relation encodes information about the evolution of the chemical enrichment
of galaxies, which in turn depends on the formation history of galaxies, their stellar mass growth, the 
feedback mechanisms regulating the star-formation processes, and the merging histories of galaxies.

The narrow-band technique to find emission lines in high redshift galaxies has been extensively applied
to large samples at different redshifts. The HiZELS survey \citep{geach08} has produced a sample of high 
redshift star-forming galaxies large enough to measure the luminosity function and clustering of \ha\ and \oii\ emitters
over the redshift range $0.4<z<2.2$ \citep{sobral10,geach12,hayashi13,sobral13}. More recently, 
using slitless grism spectroscopy from the Hubble Space Telescope, the PEARS survey sampled a large 
number of  \ha, \oii\ and \oiii\ emission line galaxies (ELGs) over the redshift range $0<z<1.5$ \citep{xia11,pirzkal13}.

At high redshifts, hydrogen recombination and forbidden lines are also targeted to derive galaxy redshifts and 
also physical properties, such as the SFR and the
gas metallicitiy { \citep[e.g.][]{maiolino08,maier06}}. Galaxies at high redshifts have been found to have lower 
gas metallicities for a given stellar mass compared
to their $z=0$ counterparts, thus suggesting an evolution of the mass-metallicity relation 
\citep{erb06,maiolino08,yuan13}. More recently, a fundamental relation between gas metallicity, stellar 
mass and star-formation rate has been suggested \citep{mannucci10,lara-lopez10} which seems to hold independent of 
redshift \citep[see, however,][]{troncoso14}. 

From a cosmological perspective, the search for the most distant galaxies is key to understand the
sources that re-ionised the Universe \citep{meiksin09,bunker10,raicevic11}.
Searching for the Lyman break {\citep{steidel96}} is a common technique to select galaxy candidates for these extreme high redshift galaxies. Normally, spectroscopic confirmation of their redshift is needed, which consists in detecting a strong emission line, such as \lya\ \citep[e.g.][]{pentericci07,stark10,schenker12}.
Emission lines are also targeted in large area surveys to measure the large scale structure of the Universe at high 
redshifts. Cosmological probes such as the detection of baryonic acoustic oscillations (BAOs) and the growth rate of cosmic structures at different 
redshifts are powerful tests that could revolutionise our cosmological paradigm. 
Ambitious programs have been proposed, including space missions such 
as EUCLID, whose goal is to detect $\sim 10^8$ \ha\ emitters in the redshift range $0.5<z<2.2$, \citep{laurejis10}, 
ground-based programs such as HETDEX \citep{hill08,blanc11}, which expects to detect $\sim 10^6$ \lya\ emitters over the redshift range $2.5<z<3.5$, 
and multi narrow-band wide field surveys such as PAU and J-PAS which have also the potential to study the large scale structure of the 
Universe traced by emission line galaxies { \citep{abramo12,benitez14}}.
 
However, despite the significant observational progress which has increased the depth and area in surveys of ELGs, 
little theoretical work has been made aiming at understanding the emission line galaxy population 
as a whole. Emission from the $^{12}$CO molecule has been studied by \citet{lagos12} in a  galaxy formation model, where
the molecular gas content is key to predict the luminosity from different rotational transitions of $^{12}$CO.
Also, a number of studies have focused in the \lya\ line, which, unlike the lines generated by the 
$^{12}$CO molecule, is originated in HII regions 
\citep[e.g.][]{ledelliou06,dayal10a,kobayashi10,orsi12}.
The intrinsic production of \lya\ photons from recombination of 
photoionised hydrogen is not a challenge to the models, 
but the high \lya\ cross section of scattering with hydrogen atoms results in the path lengths of photons being large enough to 
make them easily absorbed by even small amounts of dust. Predictions for \lya\ emitters were presented in 
detail in \citet{orsi12} including the escape of \lya\ photons through galactic outflows computed with a Monte 
Carlo radiative transfer model. 

In this work we extend the modelling of ELGs by focusing in other nebular emission lines in the optical and far-infrared (FIR) range.
The ELG population is studied by making use of a fully fledged 
semi-analytical model of galaxy formation. The \sag\ model \citep[acronym for Semi-Analytic galaxies;][]{cora06,lagos08} 
makes use of N-body dark matter simulations to follow the 
abundances and merging history of dark matter haloes where astrophysical processes shape the formation and evolution 
of galaxies. Predictions of \ha\ emission in star-forming galaxies were presented in \citet{orsi10} making use of the semi-analytical model \galform.
Here we extend the range of predictions by combining the \sag\ model with the photoionising code \mapp\ \citep{dopita00,levesque10}
to predict the luminosities of forbidden lines in the optical such as \oii\ and \oiii\ and far-infrared (FIR) 
lines such as \nii.

The paper is organised as follows: Section \eqref{sec.models} briefly describes the semi-analytical model \sag\ and the photoionisation code \mapp\ that we use in this work. Section \eqref{sec.elgmodel} presents our strategy to compute the emission lines of galaxies. 
Section \eqref{sec.results} shows how our predictions compare to observational measurements and describes predictions of the properties 
of ELGs. In Section \eqref{FIR} we present predictions for very high redshift
galaxies that could be observed with submillimetre facilities. Then, in Section \eqref{sec.discussion} we discuss the validity of our modelling 
strategy. Finally, in Section \eqref{sec.conclusions} we summarise and present the main
conclusions of this work.

\section{The models}
\label{sec.models}

We compute nebular emission from star-forming activity in galaxies by combining two different models. 
The semi-analytical model of galaxy formation \sag\ is the backbone of this work. This model predicts the physical 
and observational properties of a large number of galaxies at virtually any redshift. 
The predicted galaxy properties are combined with the \mapp\ photoionisation code to predict emission 
line luminosities due to star-forming activity. In the following, we give a brief description of both models 
and our strategy to obtain the nebular emission of star forming galaxies.

\subsection{The galaxy formation model}

\sag\ is a semi-analytical model in which the formation and evolution of galaxies 
is followed based on a hierarchical structure formation cosmology. 
In short, the model determines 
the heating and radiative cooling of gas inside 
dark matter haloes,
the formation of stars from cold gas,
the feedback mechanisms from supernova explosions and AGN activity
regulating the star formation process, 
the formation of the bulge component during
bursts of star-formation associated
to galaxy mergers or disk instabilities,
the chemical enrichment of the ISM of galaxies and 
the spectral evolution of the stellar components. 

In this work, we make use of the latest variant of \sag\ presented in \citet{gargiulo14}. % REF NEEDED FROM ARXIV
We refer the reader looking for a full description of the \sag\ model 
to \citet{cora06,lagos08,tecce10,ruiz13} and \citet{gargiulo14}. 
In the following, we describe the features of \sag\ 
that are more relevant to this work.

The backbone of the \sag\ model are merger trees of dark matter 
subhaloes taken from a N-body simulation. 
This simulation is based on the standard $\Lambda$CDM scenario,
characterised by
the cosmological parameters $\Omega_{\rm m}=0.28$,
$\Omega_{\rm b}=0.046$, $\Omega_{\Lambda}=0.72$, $h=0.7$, $n=0.96$,
$\sigma_{8}=0.82$, according
to the WMAP7 cosmology \citep{jarosik11}. The simulation was
run using GADGET-2 \citep{springel05b}
using $640^3$ particles in a cubic box of comoving side-length
$L=150\,h^{-1}{\rm Mpc}$.
In a post-processing step of the outputs of the simulation, 
the formation, evolution, merging histories and 
internal structure of dark matter haloes and subhaloes are computed by using a
`friends-of-friends' (FoF) and then the
 SUBFIND algorithms \citep{springel01}.  

{ 
Galaxies are assumed
to form in the centre of dark matter subhaloes. 
Initially, a gaseous disc with an exponential density profile is formed
from inflows of gas generated by radiative gas cooling 
of the hot gas in the halo.
When the mass of the disc cold gas exceeds a critical mass, 
an event of quiescent star formation takes place.
By assuming an initial mass function (IMF) and a certain set of stellar yields,
we can estimate the amount of metals contributed by stars in different mass 
ranges, taking into account their lifetimes.
The recycling process is the result of stellar mass loss and supernova
explosions contributing to the gradual chemical enrichment of the cold gas phase,
from which new generations of stars are formed.
The energetic feedback from core collapse supernovae produces outflows of
material that transfer the reheated cold gas with its corresponding fraction 
of metals to the hot gas phase,
being available for further gas cooling. Since the
gas cooling rate depends on the hot gas metallicity, 
the former is modified as a result of
these outflows that contribute to the chemical pollution of the hot gas.
This circulation of metals among the different baryonic components
results in the metallicity evolution of the gas in galaxies.

The cold gas of central galaxies, defined as the galaxy hosted
by the most massive subhalo
within a FoF halo,
can be replenished by infall of
cooling gas from the intergalactic medium.

Satellite galaxies are those galaxies residing in smaller subhaloes 
of the same FoF halo 
and those that have lost their subhaloes by the action of tidal forces.
The latter type of satellites eventually merge with the
central galaxy of its host subhaloes after a dynamical friction time-scale
\citep{binney87}.
Thus, 
central galaxies and satellites that keep their DM substructure 
can continue accreting stars and gas from merging
satellites.

When a galaxy becomes a satellite, it is affected by strangulation, that is, 
all of its hot gas halo is
removed and transferred to the hot gas component of the corresponding central 
galaxy and, consequently, gas cooling is suppressed.
Gas cooling in central galaxies is reduced or even suppressed 
by feedback from AGN generated as a consequence 
of the growth of supermassive black holes in galaxy centres.

%Mergers and disc instabilities trigger starbursts which contribute
%to the formation of a bulge component.  

The only channel for bulge formation in our model is via bursts of star formation, 
which occur in both mergers and triggered disc instabilities.  
In the case of major mergers, the disc stars of the progenitors are 
transferred to the bulge of the descendant.  In minor mergers, only the stars 
of the smaller progenitor feed the descendant bulge.  Discs form and grow only 
via quiescent star formation.  In this version of the model we only follow the 
macroscopic properties of these two components and cannot make direct morphological 
comparisons to clumpy, irregular high redshift galaxies.  However, we find good agreement 
between the sizes of our model galaxies and those observed in the low and high redshift universe 
\citep{padilla13}.

}

We have introduced a calculation of the spectral energy distribution (SED) of each model galaxy as a 
result of their individual star formation history, based on the stellar population synthesis code
Charlot \& Bruzual 2007, an updated version of the \citet{bruzual03} code. { In this code, the evolution 
of thermally pulsating asymptotic giant branch (TP-AGB) stars is included with
the prescription of \citet{marigo07} and \citet{marigo08}.
In these new models, young and intermediate-age stellar populations have 
younger ages and lower masses than their 2003 counterpart. We refer the 
reader to \citet{bruzual07} for more details of the models. 
}

Unlike previous versions of \sag\ where photometric data
was available for only a handful of filters at the rest-frame, the detailed calculation of the SED of galaxies
allows us to obtain magnitudes from any filter and the continuum flux around the emission lines we are interested, which is used
to compute the equivalent width (EW) of emission lines.

Early versions of \sag\ assumed that during a starburst episode all the cold gas available is converted
into stars instantaneously. This crude approximation has a dramatic impact on the ionisation photon
budget coming from galactic bulges, since ionising photons only trace SF episodes over the last $\sim$ 10 Myr. 
Hence, an instantaneous SF episode creates an unrealistic old population of bulge stars, with few or no ionising
photons.

In the latest variant of \sag, \citet{gargiulo14} introduced an extended
SF period for starbursts, in which these are characterised by a certain time-scale in which the cold gas
is gradually consumed.
The cold gas that will be eventually converted in bulge stars
is referred to as bulge cold gas, in order to differentiate it from the
disk cold gas. The time-scale for consuming the bulge cold gas
is chosen to be the dynamical time-scale of the disk. However, as the 
starburst progresses, 
effects of supernovae feedback, recycling of gas from dying stars
and black hole growth modify 
the reservoir of cold gas of both disk and bulge, thus also changing the time-scale of the starburst, 
as shown in \citet{gargiulo14}.

{
Regarding the chemical enrichment,
we follow the production of chemical elements
generated by stars
in different
mass ranges, from low- and intermediate-mass stars (LIMs) to quasi massive
and massive stars. 
The yields of \citet{karakas10} are adopted for the former 
(mass interval $1-8 M_{\odot}$). Stars in the latter mass range are
progenitors of core collapse supernovae. 
Yields resulting from mass loss of  pre-supernova stars
and explosive nucleosynthesis are taken from 
\cite{hirschi05} 
and\citet{kobayashi06}, respectively. This combination of stellar yields
are in accordance with the large number of constraints for the Milky Way
\citep{romano10}.
Ejecta from supernovae type Ia are also included, for which
we consider the nucleosynthesis
prescriptions from the updated model W7 by \citet{iwamoto99}.
The SNe Ia rates are estimated using the single degenerate model
\citep{greggio83, lia02}.
We take into account
Stellar lifetime given by \citet{padovani93} are used 
to estimate the return time-scale of mass losses and ejecta from all 
sources considered.
}

The \sag\ model contains a number of free parameters. To find an optimal set of parameter values we use the Particle Swarm Optimisation technique described in \citet{ruiz13}. The multidimensional space defined by the free parameters is explored in order to localise the minimum that reproduces a set of observables, including the local optical luminosity functions, the black hole 
mass to bulge mass relation and supernovae rates up to $z\sim 1.5$.

For simplicity, in this work we assume that stars are formed following a universal Initial Mass Function (IMF) given by 
\citet{salpeter55}. The calibration of \sag\ using the observational data described above assuming a 
Salpeter IMF can be found in the Appendix of \citet{gargiulo14}.

\subsection{The photoionisation code }
{
\mapp\ is a one dimensional shock and photoionisation code for modelling nebula emission. A detailed description can be found in \citet{dopita95,dopita96} and \citet{groves04a}, but we briefly describe the main points of the code here.

A \mapp\ photoionisation model for a simple H{\sc ii} region consists of a central point source emitting an ionising radiation field 
surrounded by a shell of constant density gas around an empty region (wind blown bubble). At each step (of constant optical depth) through the gas the code determines the ionisation and temperature balance determined from the incident ionising continuum, taking account of absorption and geometric dilution. The code includes a robust calculation of hydrogenic recombination, collisional excitation and nebular continuum processes over all wavelength, as well as a rigorous modelling of dust including absorption, charging and photoelectric heating \citep{groves04a}. At the given end of the model, typically the Str\"{o}mgen radius, the spectrum of the H{\sc ii} region including emission-line intensities, is integrated over the full ionised volume.The input parameters for such models are: the incident flux and shape of the ionising radiation field, and the metallicity and density of the ionised gas.

We specifically use in this work the precomputed H{\sc ii} region model grid of \citet{levesque10}. The incident ionisation spectra in these models are determined using the stellar population synthesis code Starburst99 \citep{leitherer99}, which predicts the ionising radiation of a simple stellar population  of a given metallicity, age and mass. The \citet{levesque10} model grid assumes large stellar clusters in the Starburst99 to fully sample the massive stellar initial mass function, and associates directly the stellar and gas-phase metallicity. The parameters and ranges of the \citet{levesque10} model grid are: the age of the stellar cluster providing the ionising radiation ($ 0 < t_{*} < 6\,{\rm Myr}$),  the metallicity of stars and gas ($0.001 < Z < 0.04$), density of the ionised gas, ($10 < n_{e} <100\,{cm}^{-3}$, and the ionisation parameter ($10^{7}<q<4\times10^{8}\,{\rm cm\,s^{-1}}$). The ionisation parameter is the ratio of the incident ionising photon flux $S_{\rm H^0}$ to the gas density, 

\begin{equation}
q=\frac{S_{\rm H^0}}{n}.
\label{eq.q}
\end{equation}

Within the SAG code we use the attenuation-corrected emission line lists from \citet{levesque10}, normalized to the H$\alpha$ line flux, and scale these lines for each SAG galaxy as described in the following section. 
We use the stellar age $t_{*}=0$ \citet{levesque10} models to represent all star-forming regions in the galaxies. This age is representative of the dominant ionising stellar populations in galaxies and reproduces the typical line ratios seen in galaxies \citep[see][ for further details]{levesque10}. 
%{\bf BRENT: why did we choose this instead of the continuous star forming models? I cant remember...}. 
We also assume a constant gas density for all galaxies of  $n_{e} = 10\,{\rm cm}^{-3}$. This gas density is typical of H{\sc ii} regions in our galaxy, but may not be representative of the more luminous objects in deep galaxy surveys. However, the lines we discuss in this work are only weakly sensitive to the range of densities observed in star-forming galaxies ( $n_{e} = 10-1000\,{\rm cm}^{-3}$), as these are well below the critical densities of these transitions (the fine-structure \nii\ line discussed in section \ref{FIR} is an exception). Importantly, our results and interpretation do not change significantly if we use the $n_{e} = 100\,{\rm cm}^{-3}$ models of \citet{levesque10}.

For the other parameters, $q$ and $Z$ we perform a bi-linear interpolation and restrict the possible range of values to that of the grid. The procedure to obtain these two values for any galaxy is described in the following section.
}
 
\section{Emission line luminosities of \sag\ galaxies}
\label{sec.elgmodel}
The line fluxes predicted by \mapp\ are conveniently characterised by two properties of the HII regions: 
the gas metallicity and the ionisation parameter. We hereafter call the flux of
the emission line $\lambda_j$ predicted by \mapp\ as $F(\lambda_j,q,Z)$, 
for a given ionisation parameter $q$ and metallicity $Z$.
Our semi-analytical model computes the properties of the 
cold gas from both the bulge and disk component of the galaxies, so we 
can track the metallicities of both components and use them as input to obtain the line fluxes from \mapp. 

%The local ionisation parameter $q$ is simply defined as 
%\begin{equation}
%q = \frac{S_{H^0}}{n} \ {\rm [s^{-1} {cm}]},
%\label{eq.q}
%\end{equation}
%where $S_{H^0}$ is the ionising photon flux per unit area, and $n$ is the local number density of hydrogen atoms. 

{ Since the semi-analytical model does not resolve the internal structure of galaxies, it is not straightforward 
to directly compute quantities such as the local ionisation parameter from Eq.\eqref{eq.q}.}

A key quantity related to the ionisation parameter is the hydrogen ionising photon rate $Q_{H^0}$, defined as
\begin{equation}
Q_{H^0} = \int_0^{\lambda_0} \frac{\lambda L_{\lambda}}{hc}{\rm d} \lambda,
\end{equation}
where $\lambda_0 = 912$\AA{}, $L_{\lambda}$ is the composite SED 
of a galaxy in units of ${\rm erg\ s^{-1}\ }$\AA{}$^{-1}$, 
$c$ is the velocity of light and $h$ the Planck constant. $Q_{H^0}$ is computed directly by \sag\ 
at each snapshot of 
the model, simply by integrating over the composite spectra predicted. 

By assuming case B recombination \citep{osterbrock89} we can express the luminosity of \ha\ 
in terms of $Q_{H^0}$ by
\begin{eqnarray}
L(H\alpha) &=& \frac{\alpha_{H\alpha}^{\rm eff} }{\alpha_B} h\nu_{H\alpha} Q_{H^0}, \\
 & = & 1.37 \times 10^{-12} Q_{H^0},
 \label{eq.QHA}
\end{eqnarray}
where $L(H\alpha)$ is in ${\rm [erg \ s^{-1}}]$,  $\alpha_{H\alpha}^{\rm eff}$ is the effective recombination 
coefficient at \ha, $\alpha_B$ is the case B recombination coefficient { for an electron temperature of $T_e = 10^4{\rm K}$}. It is expected that a fraction $f_{\rm esc}$ of 
the total ionising photon production escapes the galaxy without contributing to the production of emission lines. In such case, $Q_{H^0}$
is reduced by a factor $(1-f_{\rm esc})$ in Eq. \eqref{eq.QHA}. The calculation of the escape fraction of ionising photons is 
a long-standing challenge in galaxy formation models. Its value is key for inferring the contribution of star-forming galaxies to 
the recombination of the Universe at high redshifts \citep{raicevic11}. Deep observations searching
for Lyman-continuum photons from star-forming galaxies have yielded estimated values for $f_{\rm esc}$ as low as zero, 
although in general the values derived are small and within $f_{\rm esc} \lesssim 0.1$ 
\citep[e.g.][]{leitherer95,bergvall06,shapley06,siana07,iwata09}.
Thus, for simplicity, in this paper we assume that all ionising photons are absorbed by the neutral medium, contributing to the nebular emission
budget. This translates into the assumption of $f_{\rm esc} = 0$.

The ionisation parameter $q$ can be estimated observationally by measuring line ratios of the same species, such as ${\rm [OIII]/[OII]}$. 
It has been found that low metallicity galaxies tend to have larger values of $q$ than galaxies with { high} metallicities
\citep[e.g.][]{maier06,nagao06,groves10,shim13}. The reason for a decrease of $q$ with metallicity is due to both the opacity of the stellar winds, which 
absorb a greater fraction of ionising photons at high metallicities, and the scattering of photons at the stellar atmospheres. 
Both effects reduce the ionising flux available and thus the ionisation parameter \citep{dopita06b,dopita06a}.

In order to mimic the resulting relation between the ionisation parameter and the gas metallicity, we relate both 
quantities through the following power-law,{
\begin{equation}
%q(Z) = 2.8\times 10^7\left(\frac{Z_{\rm cold}}{0.012}\right)^{-1.3},
q(Z) = q_0 \left(\frac{Z_{\rm cold}}{Z_0}\right)^{-\gamma},
\label{eq.qZ}
\end{equation}
where $Z_{\rm cold}$ corresponds to the metallicity of the cold gas of the disk or bulge component, and $q_0$ corresponds
to the ionization parameter of a galaxy with cold gas metallicity $Z_0$. In the next section, we will explore model predictions with different 
values of $\gamma$.} 
For illustration, we also study results obtained using fixed values of $q$ that bracket the \mapp\ grid, $q= 10^7 {\rm [cm/s]}$ and $q= 4\times 10^8 {\rm [cm/s]}$.

Finally, $L(\lambda_j)$, the emission line luminosity of the line $\lambda_j$ is computed as
\begin{equation}
L(\lambda_j) =  1.37 \times 10^{-12} Q_{H^0} \frac{F(\lambda_j,q,Z_{\rm cold})}{F(H\alpha,q,Z_{\rm cold})}.
\label{eq.F2L}
\end{equation}

\section{Results}
\label{sec.results}
In this section we present several comparisons between our 
model predictions and observational measurements of the abundance 
of ELGs. To avoid an artificial incompleteness due to the limited halo mass resolution of the N-body simulation, 
the galaxies in \sag\ we use in our analysis are limited to having stellar masses 
$M_{\rm stellar} \geq 10^{9}{\rm [h^{-1}M_{\odot}]}$ and \ha\ luminosities {
$\lha \geq 10^{40} \lunits$. }

\begin{figure}
\centering
\includegraphics[width=8.5cm]{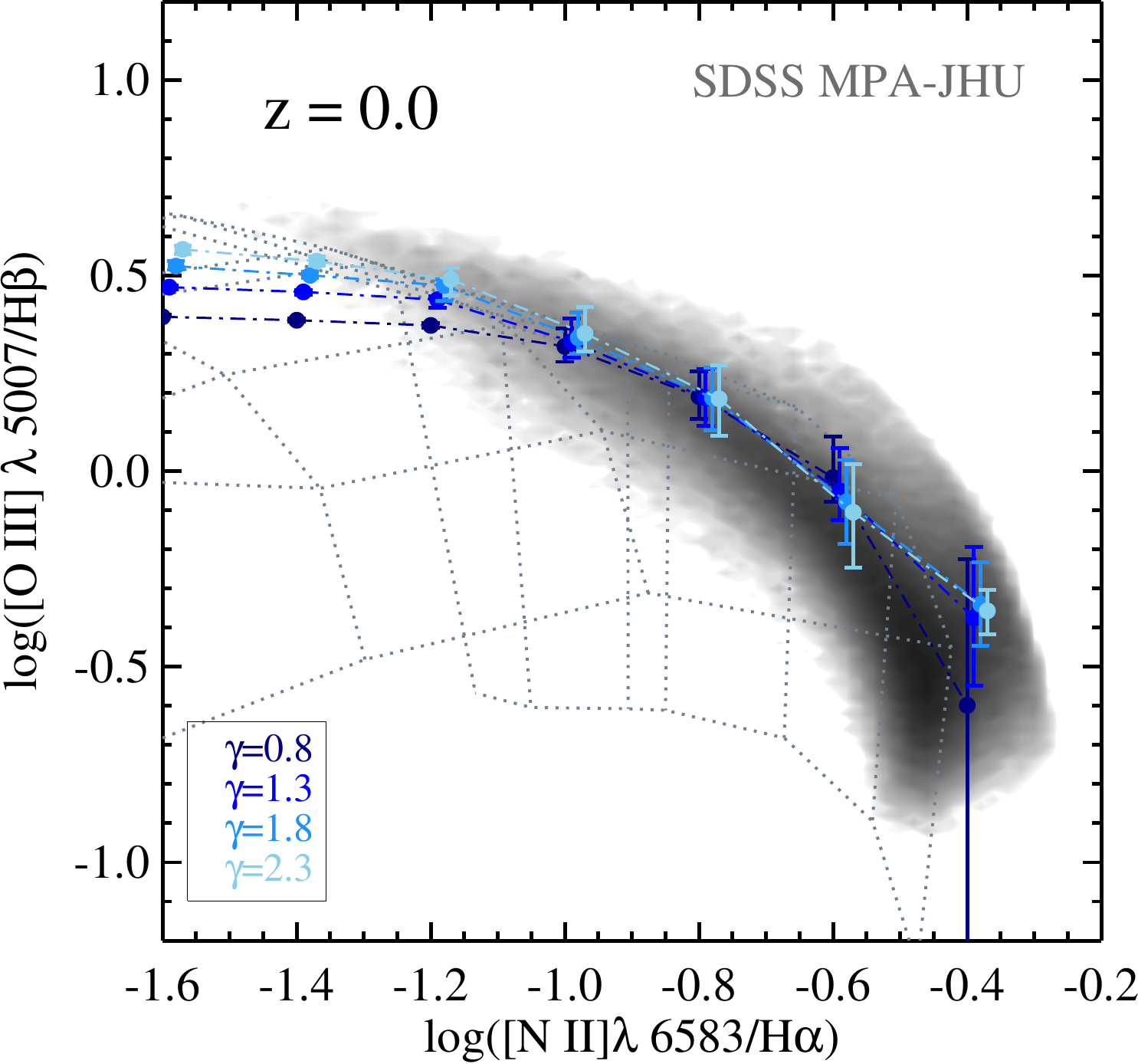}
\caption{ The BPT diagram predicted by our model (blue circles) compared with 
the emission line local galaxy sample from the SDSS MPA-JHU public catalogue (gray shaded region). Different
colour circles show
the median of the line ratios predicted by our model assuming 4 different values of the slope $\gamma$ in Eq. \eqref{eq.qZ}. The 
error bars show the 10-90 percentiles. The dotted gray lines show the line ratios predicted by \mapp\ for the set of values of $q$ and $Z$ included in the grid.}

\label{fig.bpt}
\end{figure}

\subsection{The  BPT diagram}

Line ratios are a common observed property of ELGs. 
They are used to derive the gas metallicity and ionisation parameter of both local 
and high redshift galaxies \citep{nagao06,kewley08,nakajima13,guaita13,richardson13}, to estimate 
dust attenuation corrections through Balmer line decrements \citep{nakamura04,ly07,sobral13}, 
to separate star-forming from AGN activity \citep{kauffmann03,kewley06}, 
and also to calibrate SFR indicators \citep{ly11,hayashi13}.

The so-called BPT diagram \citep{baldwin81} compares the line ratios $\oiii/{\rm H} \beta$ with 
${\rm [N II] \lambda 6584}/\ha$. This line diagnostic
is useful to distinguish between gas excited by SF (i.e. HII regions) versus that excited by AGN activity.
Qualitatively, since hydrogen recombination lines are strongly dependent on the ionising photon rate production $Q_{H^0}$, 
\ha\ and \hb\ are used in the denominators of the line ratios to remove the dependence from 
the total ionising photon budget, and also to minimise the effects of dust attenuation (because of the proximity of 
$\ha \ \lambda 6562$ to ${\rm [N II]} \lambda 6584$, and $\hb \ \lambda 4851$ to \oiii). 
Hence, both line ratios can be analysed in terms of the ionisation parameter and the metallicity of the gas.
Fig.\ref{fig.bpt} compares the line ratios of the BPT diagram for star-forming galaxies 
taken from the SDSS MPA-JHU\footnote{ http://home.strw.leidenuniv.nl/~jarle/SDSS/} 
catalogue \citep{kauffmann03} with our model predictions. 

{ In order to use our model for the ionisation parameter $q$, Eq. \eqref{eq.qZ}, we set $q_0=2.8\times10^{7} {\rm [cm/s]}$
and $Z_0 = 0.012$. This combination allows us to assign ionisation parameter values that bracket the range spanned by our \mapp\ grid
for the bulk of the galaxy population predicted by \sag. Also, we vary the slope of Eq. \eqref{eq.qZ} to lie in the range $0.8 \leq\gamma\leq 2.3$ and explore the resulting model predictions. The lower limit of $\gamma$ chosen (0.8) corresponds to the value found in \citet{dopita06b}.

Overall, our model} can reproduce the shape of the BPT 
diagram for star forming galaxies remarkably well. This is an expected result, since we are basing our 
predictions for the line ratios in a photoionisation model known to reproduce the BPT diagram if $q$ is 
chosen to be inversely proportional to the metallicity \citep{dopita06a,kewley13a,kewley13b}. {
Regardless of the value of $\gamma$ in Eq. \eqref{eq.qZ}, our model predicts line ratios that have virtually
very little scattering. Our model links the ionization parameter with the metallicity monotonically, and thus 
galaxies are predicted to have line ratios following a narrow range of values from the full grid of line ratios from \mapp, 
displayed in Fig. \ref{fig.bpt} for illustration.}

%The exact form of Eq. \eqref{eq.qZ} we propose here also allows us to predict a consistent 
%luminosity function of \oii\ and \oiii, as shown in the next section. 

{ 
When comparing the model predictions with different choices for the slope $\gamma$, the only noticeable
difference between the model is found in the upper-left corner of the BPT diagram, i.e. for high values of the ratio $\oiii/\hb$
and low values of ${\rm [NII]\lambda 6583}/\ha$. This region is characterised by high ionization parameter and low metallicities.
Hence, a steeper slope of Eq. \eqref{eq.qZ} increases the number of low-metallicity galaxies with high ionization parameter.
When comparing with the SDSS sample, the model with a steep slope ($\gamma = 2.3$) is more consistent than the others. The value
$\gamma=0.8$ determined by \citet{dopita06a} seems to be the less-favored in this case. However, it is important to notice that
in the regime of high $q$ and low $Z$ the escape fraction \fesc\ of ionising photons could depart significantly from zero \citep[see, e.g.][]{nakajima13}, 
resulting in an additional increase of the top-left region of the BPT diagram.}

A change in the shape of the BPT diagram for high redshift galaxies has recently been reported 
\citep[e.g.][]{yeh12,kewley13a,kewley13b}. This shift is probably caused by an increase of the ionisation parameter of galaxies
towards values that are outside the standard maximum allowed by the grid of configurations in \mapp\ available to us, $q=4\times 10^8 {\rm [cm/s]}$. Our model is thus unable to reproduce such evolution of the ionisation parameter, since the relation between $q$ and $Z$ 
is fixed by Eq. \eqref{eq.qZ}, making line ratios in galaxies at high redshift vary along the same sequence displayed by the model in Fig. \ref{fig.bpt}.

\subsection{The abundance of star-forming ELGs}

\begin{figure}
\centering
\includegraphics[width=8.5cm]{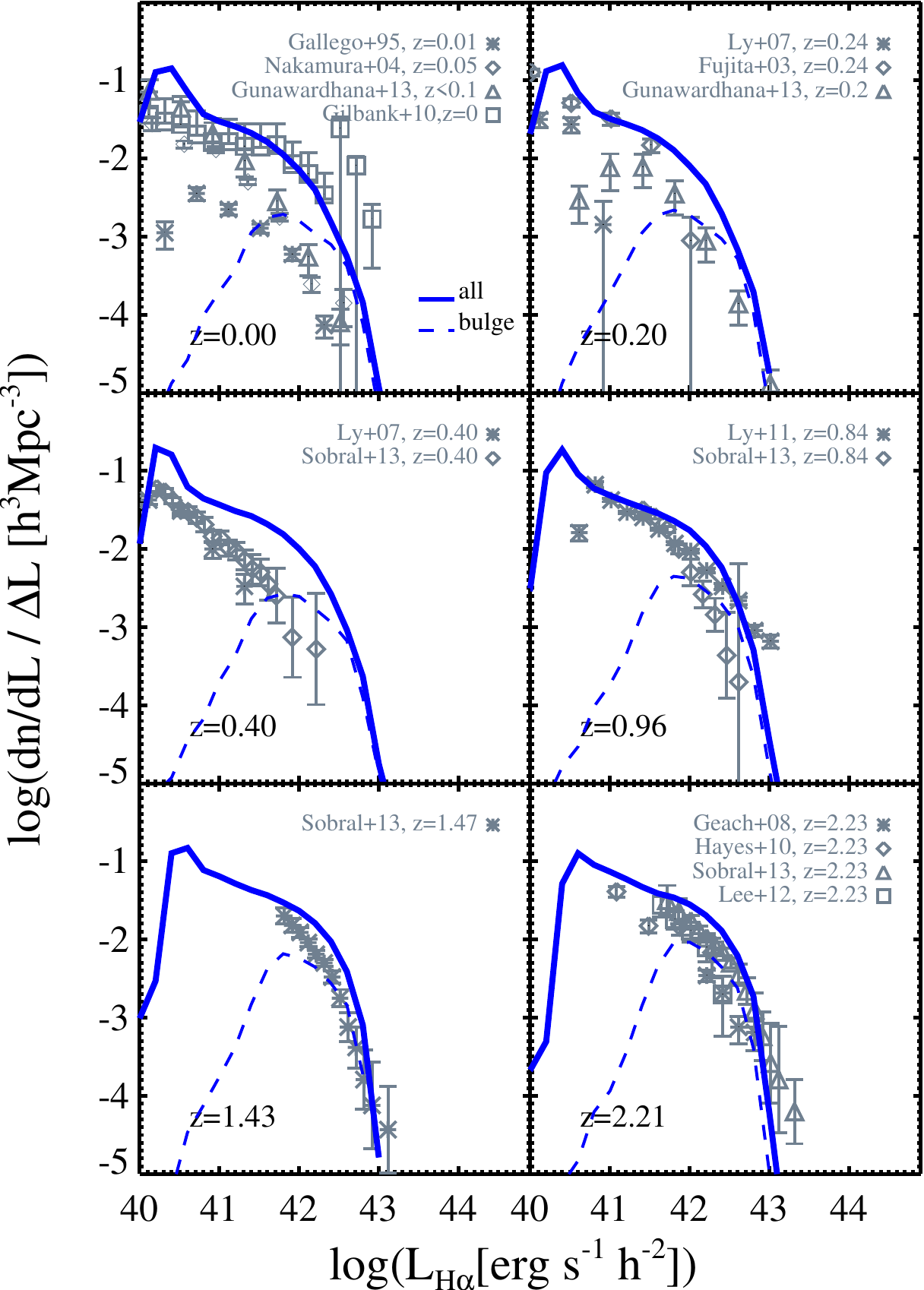}
\caption{The \ha\ luminosity functions predicted by our model  
compared to observational data at different redshifts. Model predictions are shown in blue, whereas the observational data are shown by grey symbols. The redshift of the model outputs, shown at the bottom-left side of each box, is chosen to match as closely as possible the redshift of the observations (shown in the upper-right corner of each box for each observational data set). The dashed line shows the contribution of the bulge component of galaxies to the total luminosity function.}
\label{fig.half}
\end{figure}

A basic way of characterising a galaxy population is by measuring their luminosity function (LF) and its evolution with redshift. 
Here, we compare the emission-line LF predicted by our model with observational estimates at different redshifts. 

The SFR over short timescales, of the order of $\sim 10$ Myr, is directly related with the production of ionising photons, $Q_{H^0}$, 
which is also proportional to the \ha\ luminosity through Eq. \eqref{eq.QHA}. Hence, the \ha\ luminosity correlates directly with the star-formation 
rate of galaxies in a timescale of a few Myr. A reasonable match between the predicted and observed \ha\ LF as a function of 
redshift can thus validate the predicted cosmic evolution of star formation in the \sag\ semi-analytical model.

{ In order to perform a comparison with a set of observational data, we selected those emission-line LFs that were 
corrected by dust attenuation for \ha, \oii\ and \oiii\ emitters. Also, we limit our predicted samples by an EW cut to avoid including 
galaxies that would not be observed due to their small EW. We compute the EW by simply taking the ratio between 
the line flux and the average value for the continuum within 100 \AA{} around the wavelength of the line. We choose 
a rest-frame EW$>10$\AA{}, which is consistent with the typical value for the limiting EW that is used in observations.}

\begin{figure}
\centering
\includegraphics[width=8.5cm]{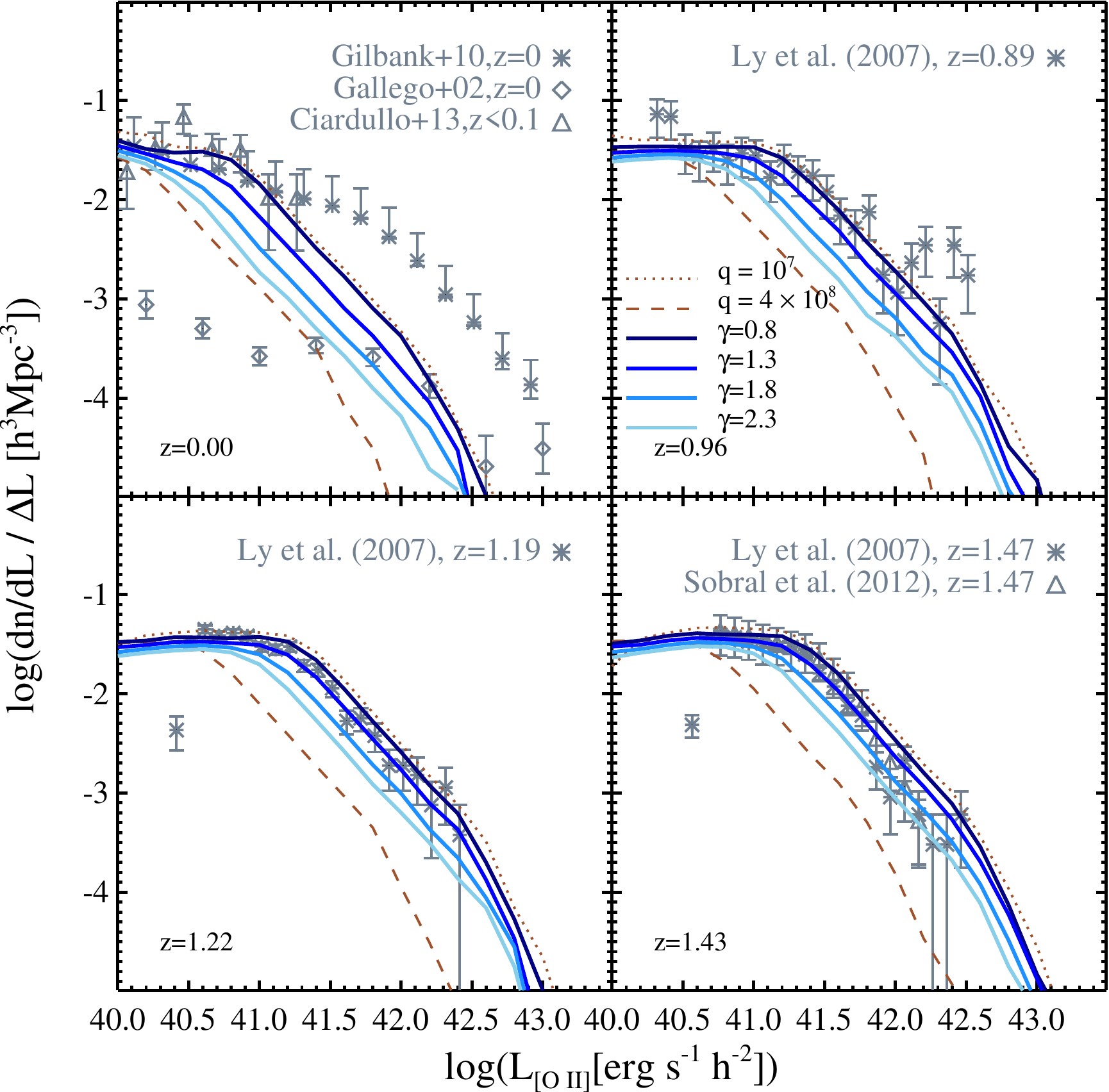}
\caption{The \oii\ luminosity functions at $z\approx 0.96$ (top panel), $z\approx 1.2$ (mid panel) and
$z\approx 1.43$ (bottom panel).
The solid blue line shows our model predictions. The solid and dashed brown lines show the results of 
assuming a constant value for the ionisation parameter of $q=10^7{\rm [cm/s]}$ and $q=4\times 10^8 {\rm [cm/s]}$
respectively. Observational data from \citet{ly07} and \citet{sobral13} are shown with grey symbols.}
\label{fig.oiilf}
\end{figure}

\begin{figure}
\centering
\includegraphics[width=8.5cm]{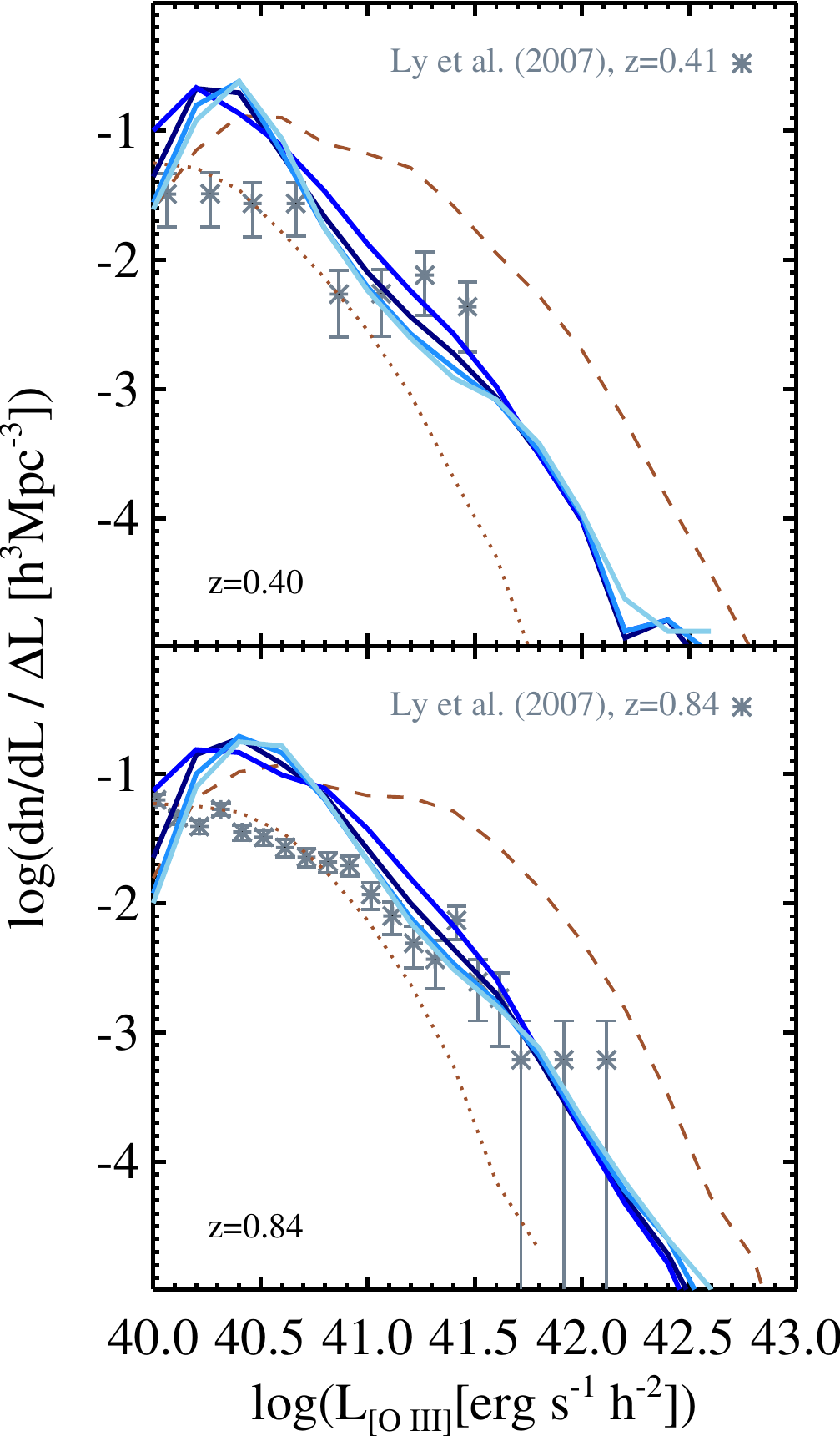}
\caption{The \oiii\ luminosity functions at $z\approx 0.4$ (top panel) and $z\approx 0.8$ (bottom panel).
The solid blue line shows our model predictions. The solid and dashed brown lines show the results of 
assuming a constant value for the ionisation parameter of $q=10^7{\rm [cm/s]}$ and $q=4\times 10^8 {\rm [cm/s]}$
respectively. Observational data from \citet{ly07} are shown with grey symbols.} 
\label{fig.oiiilf}
\end{figure}

Fig. \ref{fig.half} shows the LF of \ha\ emitters at different redshifts, predicted by the models.
Observational data is taken from {\citet{gallego95,fujita03,nakamura04,ly07,geach08,hayes10a,gilbank10,lee12,gunawardhana13} and \citet{sobral13} }
spanning the redshift range 
$0<z<2.2$. 
Overall, there is a reasonable agreement between the model predictions and the observational data, { particularly
for $z\gtrsim 0.4$.} 
At lower redshifts, our model { predictions for the \ha\ LF are consistent with some 
observational datasets, at both the faint and bright ends. 
At $z=0$, our model predictions are significantly above the observed LFs 
of \citet{nakamura04} and \citet{gallego95}, but are consistent with the faint end of the 
\citet{gunawardhana13} LF, and the \citet{gilbank10} data for a broader luminosity range.

The scatter displayed among the different observational datasets can be explained by a combination 
of different biases: cosmic variance due to different sample sizes and areas surveyed, 
stellar extinction, EW cut, ${\rm [N II]\lambda 6583}$ contamination on the total \ha\ flux and EW estimated and also 
the global dust attenuation correction applied. Most of the datasets used to compare our models with apply the 
Balmer decrement technique, in which an additional recombination line (typically \hb) is used to estimate the deviation
of the line ratio from what is expected from recombination theory, and thus estimating the dust attenuation necessary
to account for this deviation. Another less precise but still common practice is to assume an average extinction $A_{H\alpha} = 1$, employed in the LFs of \citet{fujita03} and \citet{sobral13} in Fig. \ref{fig.half}. 

An additional important caveat of the observed dust-corrected LFs is that these are likely to lack
objects that are excessively attenuated so that they were not detected despite being intrinsically 
bright enough to appear in a dust-free LF.}

{ The predicted \ha\ luminosity in our model} does not depend on the model for the ionisation 
parameter. Therefore, the predicted \ha\ LFs shown in Fig. \ref{fig.half} are valid for any model of the ionisation 
parameter. 

The excess of SFR in \sag\ implied by the \ha\ luminosities at low redshift has also been studied in \citet{ruiz13}, 
where the observed decline in the specific SFR from $z\sim 1$ towards $z=0$ is shown to be 
steeper than what is predicted by \sag. 

The dashed lines in Fig. \ref{fig.half} show the contribution of star formation occurring in the bulge to the total luminosity function. 
Overall, bursty star-formation due to either merger episodes or disc instabilities contributes significantly only for \ha\ 
luminosities $L_{H\alpha} > 10^{43} \lunits$ at any redshift. This means that only the brightest sources have a 
significant contribution of line luminosity coming from the bulge.  

In order to test our choice of the model for the ionisation parameter, we study the LF of forbidden lines. 
{ Fig. \ref{fig.oiilf} compares the model predictions for the \oii\ LF with observational datasets
from \citet{gallego02}, \citet{ly07},\citet{gilbank10}, \citet{ciardullo13} and \citet{sobral13} spanning the redshift range $0\lesssim z \lesssim 1.4$.
 
We show the predicted LFs in our model assuming constant values of $q$ at the lowest and highest possible values
within the \mapp\ grid, $q = 10^7 {\rm cm/s}$ and $q = 4\times 10^8 {\rm cm/s}$. Also, like in Fig. \ref{fig.bpt}, we
explore the result of choosing different values for the slope $\gamma$ of the $q-Z$ relation, Eq. \eqref{eq.qZ}.

The predicted \oii\ LF at $z=0$ is found to be in partial agreement with the observational data. Models with 
values of $\gamma = 0.8$ and $1.3$ are consistent with the faint end of the \citet{gilbank10} and \citet{ciardullo13}
LFs, and these are also marginally consistent with the bright end of the LF measured in \citet{gallego02}. However, 
overall the model predictions are not consistent with the full luminosity range covered by the LF of \citet{gilbank10}. 
Differences in the estimated LFs between the different datasets could be related with the luminosity correction 
due to dust attenuation. The \oii\ LFs of \citet{gilbank10} and \citet{ciardullo13} are corrected by dust attenuation
using the Balmer decrement technique, and in \citet{gallego02} dust attenuation is corrected using the 
measured colour excess $E(B-V)$.

At higher redshifts, the model predicted \oii\ LFs are in remarkable agreement with the observational measurements of 
\citet{ly07} and \citet{sobral13} in the redshift range $0.9\lesssim z \lesssim 1.4$. 
In particular, the models with a constant low ionization parameter of 
$q=10^7 {\rm [cm/s]}$, and those with $\gamma=0.8$ and $1.3$ are favored. For steeper slopes, 
the LF falls under the observed LFs. 

We also study the abundance of \oiii\ emitters at different redshifts. Fig. \ref{fig.oiiilf} shows the predicted \oiii\ luminosity functions compared to observational data from \citet{ly07} 
at redshifts $z=0.4$ and $z=0.8$.
We find an interesting trend in the predicted LFs of \oiii, when compared to the observational data. 
Both predictions using a constant ionisation parameter appear to enclose the observed LF of \oiii. The model 
predictions with low $q$ show a LF below the one when assuming a high value of $q$. The observational data
seems to lie in between, somewhat favouring low values of $q$ at bright \oiii\
luminosities. Our model predictions with $q$ defined by Eq.\eqref{eq.qZ} interpolates between these two 
regimes, being overall more consistent with the observational data than the models with a constant $q$. Unlike
the \oii\ LFs in Fig. \ref{fig.oiilf}, the predicted \oiii\ LF has a nearly negligible dependence on the 
values of the slope $\gamma$. Fig. \ref{fig.oiiilf} shows, however, that it is necessary to invoke a relation between
$q$ and $Z$ to reproduce the observed \oiii\ LFs. 

Interestingly, the \oiii\ LFs predicted with constant ionization parameters are inverted compared with the \oii\ LFs. 
The ionization parameter is roughly proportional to the line ratio $\oiii/\oii$. This makes galaxies with a high ionization 
parameter to have a brighter \oiii\ luminosity and fainter \oii\ luminosity with respect to a galaxy with a low 
ionization parameter. This explains why the case with a constant high ionization parameter predicts a brighter LF for \oiii\
and a fainter one for \oii.

From the analysis of the BPT diagram and the emission-line LFs we find that the models that best match the observational 
datasets are those with $\gamma = 0.8$ and $\gamma=1.3$. Hence, hereafter we choose to show model predictions for the model with 
$\gamma=1.3$, i.e. a model that computes the ionization parameter of galaxies by using the following relation:

\begin{equation}
q(Z) = 2.8\times 10^7\left(\frac{Z_{\rm cold}}{0.012}\right)^{-1.3}.
\label{eq.qZmodel}
\end{equation}

}

\subsection{The Star-formation rate traced by emission lines}

\begin{figure*}
\centering
\includegraphics[width=14cm]{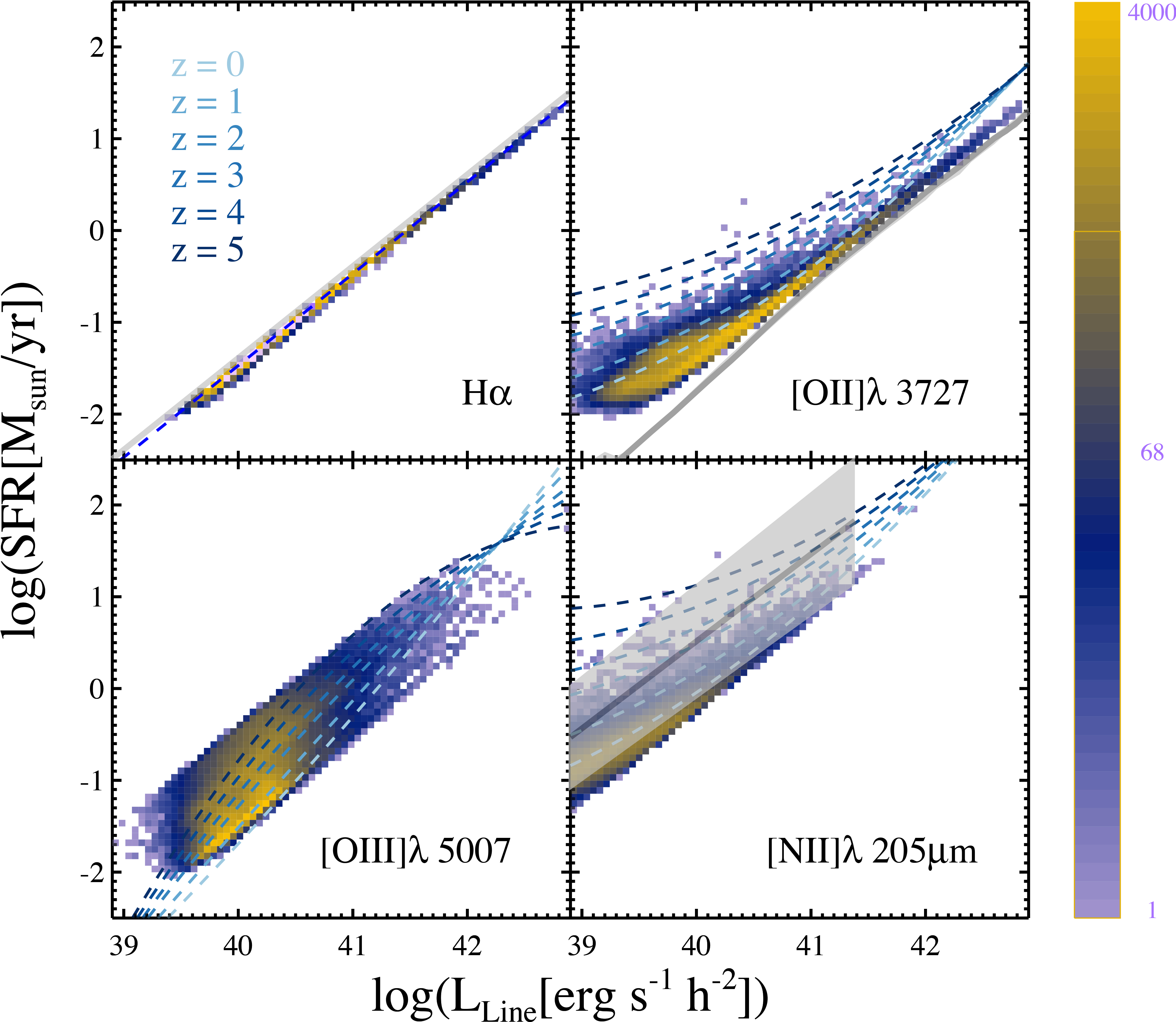}
\caption{The predicted relation between the star-formation rate and the \ha, \oii, \oiii\ and \nii\ line luminosities, 
as shown by the legend. { The coloured regions show the predicted density of galaxies in the plot for $z=0$. The blue dashed 
lines show fits to this relation for redshifts $z=0$ to $z=5$ from bright to darker. Estimates of the SFR based on observational data for different lines are shown in gray: \citet{kennicutt98b} for \ha, \citet{kewley04} for \oii\ and \citet{zhao13} for \nii\ (see text for details).}}
\label{fig.SFR}
\end{figure*}

The \ha\ luminosity is the most common tracer of star forming activity occurring within a few Myr. Its rest-frame 
wavelength ($\lambda_{H\alpha} = 6562$\AA{}) makes it accessible to optical and near-infrared facilities spanning 
the redshift range $0<z<2.2$. The inferred star-formation rate density using \ha\ as a function of redshift has 
been shown to be consistent with other star-formation estimators, such as the UV continuum 
\citep{hopkins04,sobral13}. Moreover, dust extinction is usually less severe at the \ha\ wavelength than at shorter 
wavelengths. However, a variety of nebular lines are produced in HII regions and therefore these 
also correlate with the instantaneous star formation rate in some way. High redshift surveys that have no 
access to the \ha\ line have to rely on alternative recombination lines, such as H$\beta$ (if detected) 
or use crude estimations based on SED fitting with a handful of photometric bands. This motivates the exploration
of using non-standard forbidden lines as tracers of star-formation rate.

Fig. \ref{fig.SFR} shows our model predictions for the correlation between the \ha, \oii, \oiii and \nii\ lines with the 
star-formation rate at different redshifts. 

Since the luminosity of a collisional excitation line depends on the ionisation parameter, 
which in turn depends on the cold gas metallicity, in our model the relation between the line luminosities 
and the star formation rate evolves with redshift,
reflecting the evolution of the cold gas metallicity of galaxies { with redshift}.

We find that it is possible to describe the relation between the star formation rate and a line luminosity, at a given
redshift, by a second-order polynomial. Each term of the polynomial is found to evolve linearly with redshift, 
thus leading to a polynomial of the form

\begin{table}
\caption{Values of the constants in Eq. \eqref{eq.sfr_fit} defining the relation between the emission line 
luminosities and the star-formation rate. Each row corresponds to a different emission line.}
\label{table.sfr_fit}
\begin{tabular}{@{}lccccccc}
\hline
\hline
Line & $ a$ & $b$ & $c$ & $d$ & $e$ & $f$\\
\hline
\hline
\ha & \aHa & - & - & - & 1.00 & - \\
\hline
 \oii & \aOII & \bOII & \cOII & \dOII & \eOII & \fOII \\
\hline
\oiii & \aOIII & \bOIII & \cOIII & \dOIII & \eOIII & \fOIII \\
\hline
\nii & \aNII & \bNII & \cNII & \dNII & \eNII & \fNII \\
\hline
\hline
\end{tabular}
\end{table}

\begin{eqnarray}
\log \dot{M}(L_{\lambda},z) & = & a + (1+z)\left[ b + \log L_{\lambda}(c + d\log L_{\lambda}) \right] + \nonumber \\
 & & \log L_{\lambda} (e + f \log L_{\lambda}),
\label{eq.sfr_fit}
\end{eqnarray}
where $\dot{M} = \dot{M}{\rm [M_{\odot}yr^{-1}]}$ is the instantaneous star-formation rate, 
$L_{\lambda} = L_{\lambda} \lunits$ corresponds to the luminosity of the line $\lambda$,
and the constants $a, b,c,d,e$ and $f$ are chosen by minimising $\chi^2$ using Eq. \eqref{eq.sfr_fit} and are 
different for each line. Table \ref{table.sfr_fit} shows the values of the constants for each line, and the dashed lines 
in Fig. \ref{fig.SFR} shows Eq. \eqref{eq.sfr_fit} and its evolution from $z=5$ down to $z=0$. 

As expected, the \ha\ line shows a tight correlation with the SFR. In our model, there is a direct relation 
between the production rate of ionising photons, $Q_{H^0}$, and the \ha\ luminosity, Eq. 
\eqref{eq.QHA}, and the latter correlates tightly with the instantaneous star-formation rate.
This relation does not evolve with redshift since both the \ha\ luminosity and the star formation 
rate are related to each other through $Q_{H^0}$ only. The ionisation parameter makes no significant
difference in our predictions for recombination lines.

{ Fig. \ref{fig.SFR} also shows a good agreement with the SFR estimate from \citet{kennicutt98b}
\citep[see also][]{kennicutt12}. Our model predicts 
\ha\ luminosities that are systematically slightly brighter for a given SFR, which is related to our assumption of 
zero-age stellar populations driving the production of emission lines.}

The \oii\ luminosity correlates strongly with the SFR, although with a significant 
scatter. Interestingly, the scatter is reduced, from about an order of magnitude at faint luminosities of $L$ 
$\sim 10^{39} {\rm [erg s^{-1} h^{-2}]}$ to virtually no scatter for  $L \gtrsim 10^{42} {\rm [erg s^{-1} h^{-2}]}$.

Also, our model predicts that a fixed \oii\ luminosity corresponds to higher SFRs as we look towards higher 
redshifts. For instance, at $L = 10^{41} \lunits$, the SFR at $z=5$ is $\sim 1 {\rm [M_{\odot}/yr]}$, and
$\sim 0.1 {\rm [M_{\odot}/yr]}$ at $z=0$.

{ We compare our relation between \oii\ and the SFR with the estimates from \citet{kewley04} (their Eq. 10). In order 
to take into account the effect of metal abundances in the production of \oii, \citet{kewley04} constructs a relation
that, starting from the \citet{kennicutt98b} formula, includes the oxygen abundance ${\rm [O/H]}$ in the relation. 
Hence, to compare their results
with our model predictions we calculate the $10-90$ percentile of ${\rm [O/H]}$ for different \oii\ luminosity bins. The resulting
scatter is, however, small, since the \citet{kewley04} relation is only weakly depending on the oxygen abundance.

Overall, the \citet{kewley04} relation obtains SFRs that are below our model predictions for a given \oii\ luminosity.
For luminosities $L(\oii)<10^{40} \lunits$, the SFR predicted is nearly an order o magnitude below our model predictions. However, for $L(\oii)\geq 10^{41} \lunits$, the \citet{kewley04} estimate is in reasonable agreement with our model predictions.}

The \oiii\ vs. SFR relation presents a large amount of scatter, regardless of the line luminosity. Overall, 
the polynomial fit of this relation shows that at a fixed \oiii\ luminosity the SFR increases towards higher
redshifts. However, it is clear from Fig. \ref{fig.SFR} that \oiii\ is the less favored SFR indicator from the ones
studied here. At around the wavelength of \oiii, it is more convenient to use the \hb\ line ($\lambda_{H\beta} = 4861$\AA{}) 
as a SFR estimator, since \hb\ is not sensitive to the ionisation 
parameter or metallicity of the ISM.

The \nii\ far infrared line is an interesting line to study since it can be detected locally with FIR space based 
facilities such as Herschel \citep{zhao13}, and at high redshifts with submillimetre facilities such as the IRAM 
Plateau de Bure Interferometer \citep{decarli12} and ALMA \citep{nagao12}.
Fig. \ref{fig.SFR} shows that faint \nii\ luminosities present a scatter of about an order of magnitude in SFR. 
However, this scatter is significantly reduced for brighter luminosities. The median SFR for a fixed \nii\ 
luminosity tends to increase for higher redshifts.

{ We compare our model predictions with estimates from \citet{zhao13}, where \nii\ emission is measured
from a sample of FIR galaxies, and the SFR is estimated from the total FIR flux using the relation in \citet{kennicutt12}, and 
then relating this with the measured \nii\ flux.  Within their measured scatter, the \citet{zhao13} is
consistent with our model predictions.}

\begin{figure}
\centering
\includegraphics[width=8.5cm]{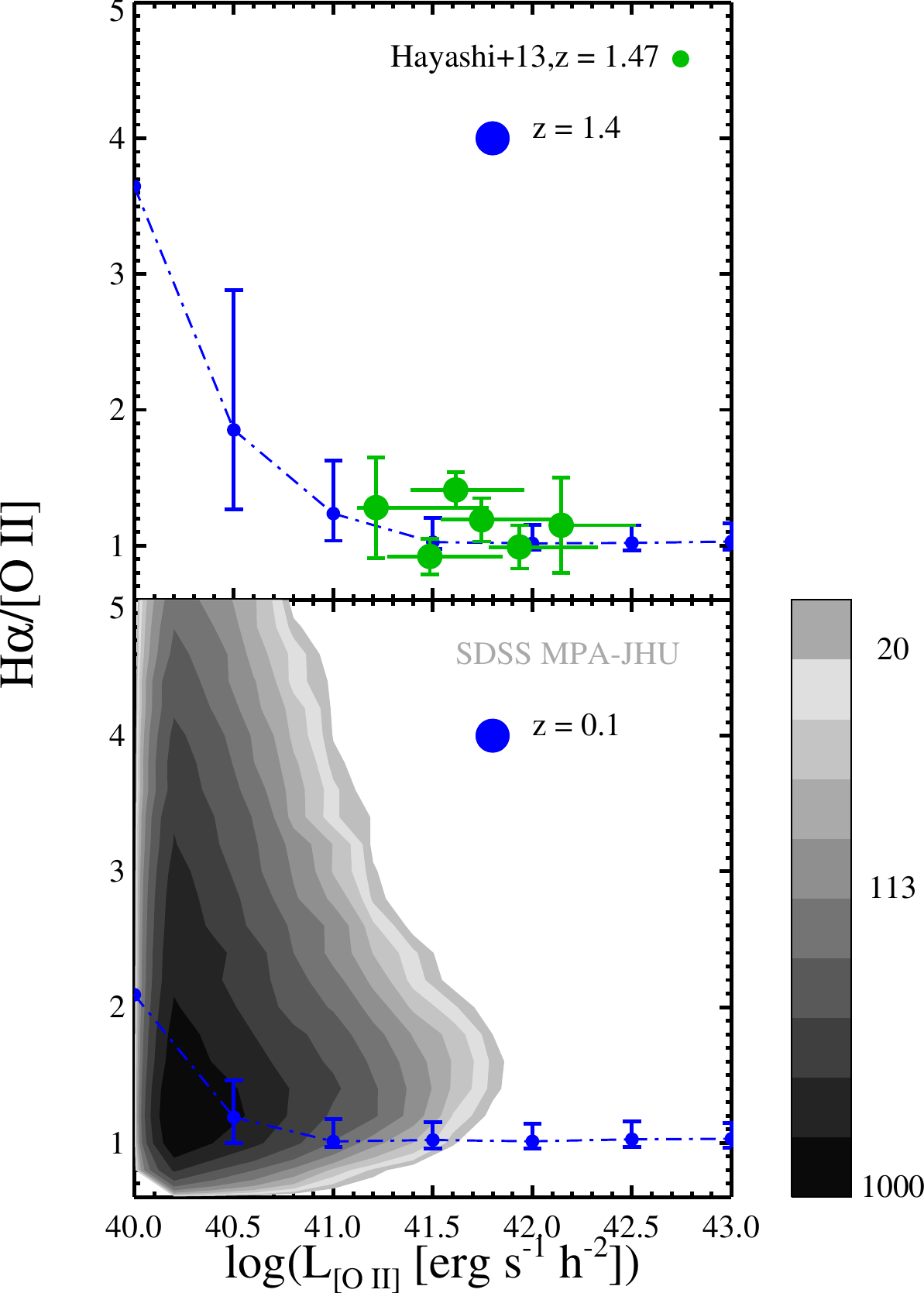}
\caption{The ratio of \ha\ to \oii\ at redshifts $z=1.47$ (top) and $z=0$ (bottom). The median of the
model predictions are shown by the blue circles. The error bars represent the 10-90 percentile of the distribution
at each \oii\ luminosity bin. Observational data from \citet{hayashi13} is shown at $z=1.47$ (green circles), and from the SDSS 
\citep{kauffmann03} at { $z \sim 0.1$ displayed as contours in gray scale}.}
\label{fig.haoii}
\end{figure}

The model predictions shown in Fig. \ref{fig.SFR} are not straightforward to validate. High redshift surveys do not typically attempt to measure the 
SFR using forbidden lines, since the dependence of the line luminosity with the properties of the HII other than the ionising photon rate is complicated.
Even our simple modelling displays a large scatter in the star formation rates that galaxies with a given line luminosity can have. Despite this, the 
scaling relations using Eq. \eqref{eq.sfr_fit} and Table \ref{table.sfr_fit} can provide an approximate value to the star-formation rate of galaxies within
an order of magnitude or less.

\citet{hayashi13} measured the 
ratio of \ha\ to \oii\ in a sample of line emitters at $z=1.47$, in an attempt to calibrate \oii\ as a star-formation rate estimator. We compare
our model predictions with the observed line ratios to validate the line luminosities our model predicts. Fig. \ref{fig.haoii} shows 
a comparison between the \citet{hayashi13} \ha\ to \oii\ ratio of star-forming galaxies at $z=1.47$ and our model predictions. Also, we 
show $\ha/{\rm [O II]}$ at $z=0$ from SDSS.

\begin{figure}
\centering
\includegraphics[width=8.5cm]{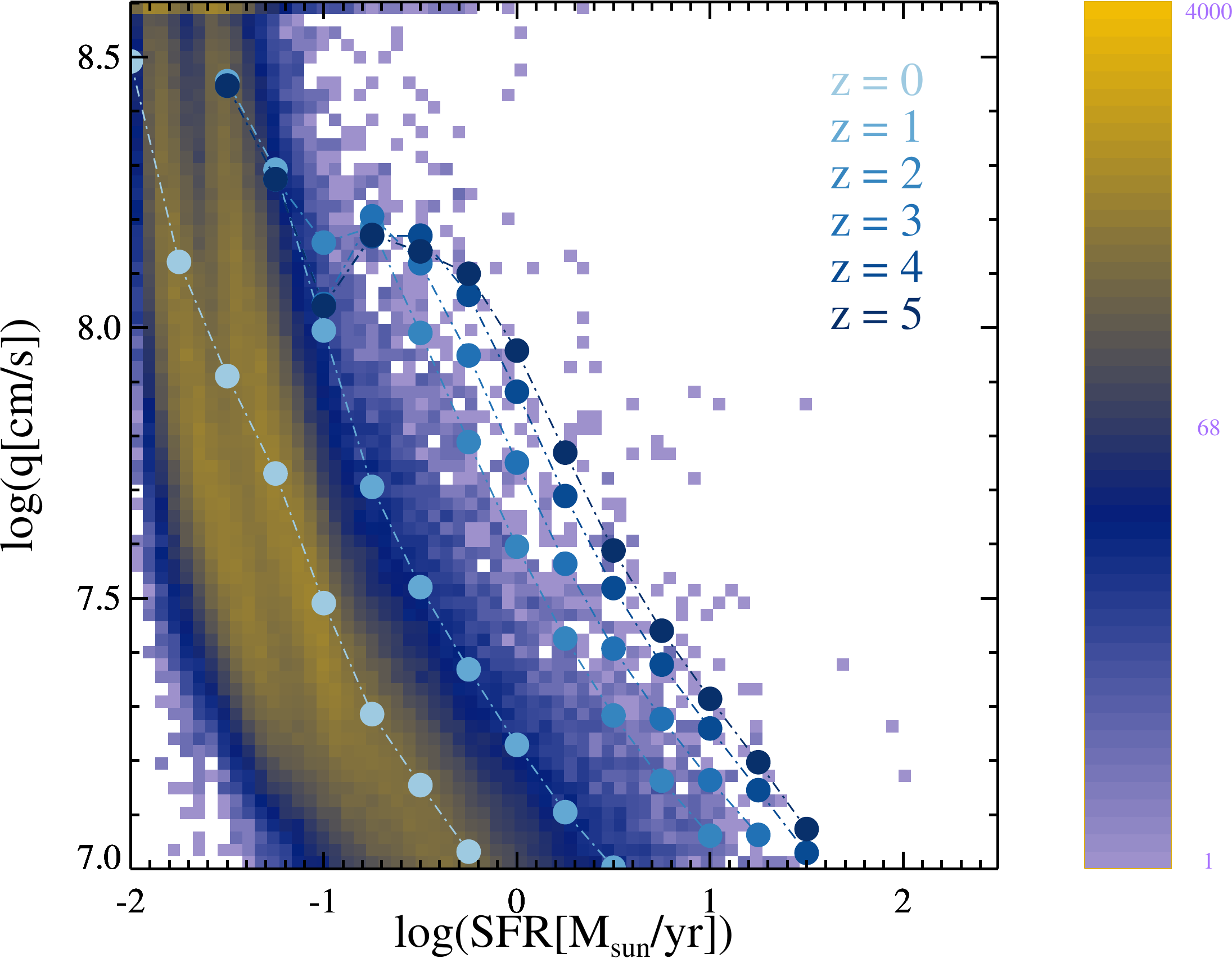}
\caption{The ionisation parameter as a function of star formation rate for redshifts $0<z<5$. The squares show the 
position of galaxies at $z=0$ for this relation, with bright yellow colours indicating a high concentration of galaxies, and 
dark blue colors showing less galaxies. Solid circles show the median ionisation parameter value at different redshifts.}
\label{fig.QSFR}
\end{figure}

Our model predicts a steep decrease in the ratio towards brighter \oii\ luminosities. At $z=1.47$, \ha\ is four times brighter than \oii\ for galaxies with
$L_{\rm [O II]} \sim 10^{40} \lunits$. At $z=0$, on the other hand, the \ha\ luminosity is twice as bright for the same \oii\ luminosity.
Interestingly, galaxies with $L_{\rm [O II]} > 10^{42} \lunits$ display a constant ratio, $\ha/{\rm [OII]} \approx 1.1$. This occurs because, as shown 
in Fig. \ref{fig.oiilf}, the majority of galaxies with bright \oii\ luminosities have a constant ionisation parameter value, $q=10^7 {\rm [cms^{-1}]}$, 
making the line ratio constant.

Fig. \ref{fig.haoii} also shows that our model predicts \ha\ to \oii\ line ratios that match those of \citet{hayashi13} at $z=1.47$ remarkably well. \citet{hayashi13}
observed line ratios are in the range where the line ratios depart from their constant value towards higher ones.
In addition, at $z\sim 0$, we show that our 
model also predicts line ratios that are consistent with the bulk of the line ratios found in the SDSS spectroscopic sample \citep{kauffmann03}.{ 
This observational sample corresponds to star-forming galaxies spanning a redshift range $z\lesssim 0.3$, although most of the galaxies are at 
$z\approx 0.1$, which is the redshift that we choose in our model to compare with.} This
comparison validates the \oii\ luminosities that are obtained with our model and their relation with the SFR.

The evolution of the relation between the SFR and the line luminosities shown in Fig. \ref{fig.SFR} arises due to an evolution of the typical 
ionisation parameter of galaxies with redshift. In our model this evolution is driven purely by the cold gas metallicity of galaxies, 
which is typically lower for high redshift galaxies. 
Fig. \ref{fig.QSFR} shows the ionisation parameter of galaxies between redshifts $0<z<5$ as a function of SFR. The median ionisation parameter
tends to be higher towards higher redshifts for a given SFR. This means that for a given SFR, galaxies have different luminosities depending on the value 
of the ionisation parameter. This creates the evolution of the SFR-line luminosity relation shown in Fig. \ref{fig.SFR}. 
Also, the scatter in this relation, illustrated at $z=0$ in Fig. \ref{fig.QSFR} is responsible for the scatter in the SFR versus line luminosity relation.

The evolution of the ionisation parameter with redshift has been suggested as a way to explain the 
apparent evolution of line ratios in high redshift galaxies \citep[e.g.][]{brinchmann08,hainline09}. By analysing different line ratios for galaxy samples within
the redshift range $z=0-3$, \citet{nakajima13} shows that at high redshift star forming galaxies seem to have typical ionisation parameter much higher than 
their low redshift counterparts. A similar conclusion has been reported by \citet{richardson13} studying high redshift \lya\ emitters. 
Both works conclude that is necessary to extend photoionisation models to include configurations with ionisation parameters greater than 
$q>10^9{\rm [cm s^{-1}]}$. However, it is important to notice that an evolution of the observed line ratios can arise by other factors than a 
sole evolution of the ionisation parameter, such as the geometry and the electron density of the gas \citep{kewley13a}.

\begin{figure*}
\centering
\includegraphics[width=16cm]{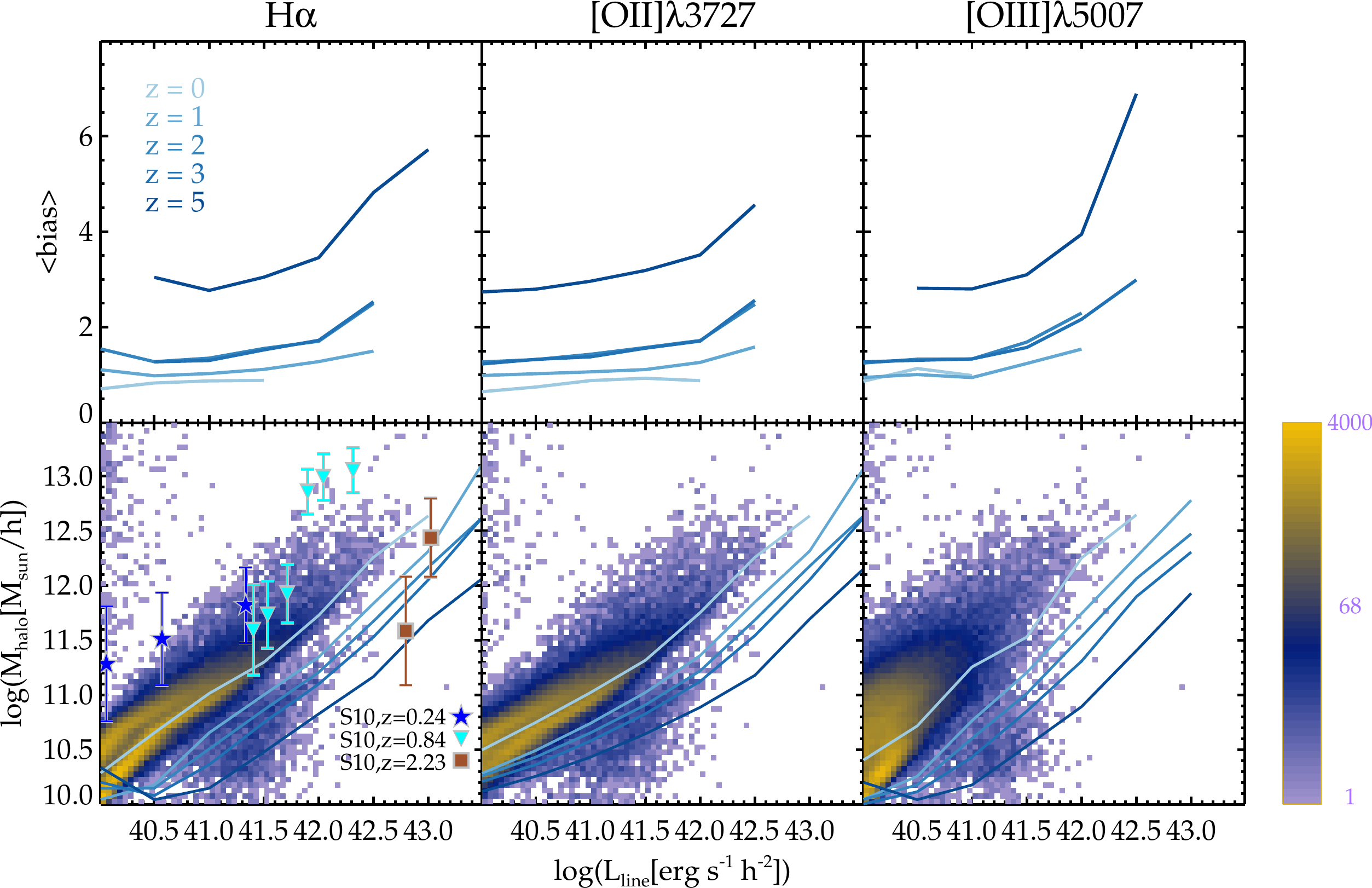}
\caption{Top panel: Galaxy linear bias for galaxy samples in the redshift range $0<z<5$ for \ha\ (left), \oii\ (middle) and \oiii\ (right) emitters. The bias is computed
for galaxy catalogues with logarithmic luminosity bins of $0.25$. 
Bottom panel: The dark matter halo mass as a function of line luminosities, for different redshifts. The coloured squares show all the range of halo masses at different luminosities, 
with brighter yellow colours indicating greater concentrations of galaxies (see the scale on the right). 
The solid lines show the median halo mass as a function of emission line luminosity for different redshifts.
Observational estimates of the minimum dark matter halo mass as a function of \ha\ luminosity from \citet{sobral10} are shown with stars, triangles and squares, for $z=0.24, 0.84$ and $z=2.23$, respectively.  }
\label{fig.bias_mhalo}
\end{figure*}

\subsection{The large scale structure traced by emission lines}

Thanks to the increasing development of wide area narrow band surveys of line emitters, it has been possible to map 
the large scale structure of the Universe at high redshifts using ELGs as tracers \citep{ouchi05,shioya08,geach08,sobral10}. 
However, little is understood in terms of how line emitters trace the dark matter structure, and how the clustering of this galaxy 
population depends on their physical properties. 

In order to quantify how the line luminosity selection of galaxies trace the underlying dark matter structure, we compute the linear bias, 
$b = (\xi_{\rm gal}/\xi_{\rm dm})^{1/2}$, where $\xi_{\rm gal}$ and $\xi_{\rm dm}$ are the two-point autocorrelation functions of 
galaxies and dark matter, respectively. Different samples of ELGs are created by splitting them in bins of luminosity with $\Delta \log L = 0.5$ 
and at different redshifts. Then, the autocorrelation function is computed for each galaxy sample. The dark matter auto-correlation 
function is computed using a diluted sample of particles from the N-body simulation at different redshifts. The dilution is done in order to compute the correlation function of dark matter efficiently but also so that the error in the bias is dominated by the sample of emission line galaxies. The adopted value for the 
linear bias is taken by averaging the ratio $\xi_{\rm gal}/\xi_{\rm dm}$ over the scales $5-15 {\rm [Mpc/h]}$, where both correlation 
functions have roughly the same shape but different normalisation.  

The top panel of Fig. \ref{fig.bias_mhalo} shows the linear bias computed for samples selected with different limiting luminosities in 
the redshift range $0<z<5$. Overall, the bias parameter increases towards higher redshifts for all line luminosities, meaning that ELGs
are increasingly tracing higher peaks of the matter density field towards high redshifts. ELGs at $z=0$ have a bias parameter 
$b \approx 1$, but at $z=5$ the bias grows rapidly towards 
$b=3-6$ depending on the line luminosity.

For a given emission line, in general a brighter sample of galaxies is predicted to have a higher bias parameter. The dependence of the bias 
parameter with the line luminosity also gets steeper towards higher redshifts. At $z=0$, there is no noticeable dependence between the bias 
and the line luminosities. At $z=1$, the bias parameter increases on average by 0.5 between $L_{\rm line} \sim 10^{40}-10^{43} \lunits$. 
At $z=5$, the bias parameter increases from roughly $b\approx 3$ to $b\approx 6$ depending on the emission line. 

The bias parameter predicted for ELGs is overall the same regardless of the emission line chosen. At $z=5$, however, bright, \ha\ emitters with $L_{H\alpha} \sim 10^{43} \lunits$ are predicted to have a bias $b \approx 6$. \oiii\ emitters
at $z=5$ are predicted to have $b \approx 7$ for $L = 10^{43} \lunits$.
\oii\ emitters, on the other hand, are predicted to reach $b \approx 4.5$ at $z=5$ for a sample with luminosity $L = 10^{42.5} \lunits$. 

At lower redshifts, emission line galaxy samples at $z=2$ and $z=3$ are predicted to have the same linear bias overall. In this case, the expected increase in the bias from $z=2$ to $z=3$ is balanced by the decrease in the typical halo masses hosting $z=3$ line emitters, which results in both populations having a remarkably similar bias.  

Observational measurements of the linear bias are normally used to infer a typical dark matter halo mass of a sample of galaxies. This is done
by making use of a cosmological evolving bias model which makes $b = b(M_{\rm halo},z)$ a function
of halo mass and redshift \citep[e.g.][]{mo96,moscardini98, sheth01,seljak04}.

The bottom panel of Fig. \ref{fig.bias_mhalo} shows the predicted dark matter halo mass of line emitters at different luminosities and redshifts, taken directly from our model. The median dark matter halo mass is found to correlate strongly with the luminosity of the samples. As an illustration, we show the
full scatter of this relation at $z=0$, displaying the scatter around the median value of halo mass that can range up to 1 order of magnitude or more. 
For a fixed luminosity, the median halo mass is larger for decreasing 
redshift. Throughout the redshift range $0<z<5$, \ha\ and \oii\ emitters span the halo mass range $10<\log(M_{\rm halo}{\rm [h^{-1}M_{\odot}]} < 12$ 
for luminosities between $10^{40}<L_{\rm H\alpha}\lunits<10^{43}$. \oiii\ is predicted to have a steep relation between
line luminosity and halo masses. This explains the higher clustering bias of \oiii\ compared to the other lines. For a given line luminosity, 
the typical halo mass of \oiii\ is in general higher than the typical halo mass obtained for the same line luminosity with \ha\ or \oii.

We compare our predictions of the dark matter haloes of emission line galaxies with available observational data. \citet{sobral10} performed 
a clustering analysis of a sample of $z=0.84$ \ha\ emitters and compared it with \ha\ samples at $z=0.24$ and $z=2.23$. Their derived 
typical halo mass for samples of \ha\ emitters of a given luminosity are shown in Fig. \ref{fig.bias_mhalo}. These were computed 
by estimating the real-space autocorrelation function of \ha\ emitters and making use of the \citet{moscardini98} evolving bias model to 
estimate the minimum halo mass of each \ha\ sample.

For simplicity, we compare our $z=0$ predictions with the observational sample of \ha\ emitters at $z=0.24$, the $z=0.84$ observational sample with 
our $z=1$ predictions, and the $z=2.23$ observational sample with our predicted $z=2$ sample. Our model predictions are
shown to be marginally consistent with the estimated \ha-halo
mass relation at $z=0.24$, and fairly consistent with the high redshift sample at $z=2.23$. However, the observationally derived halo masses
of \ha\ emitters at $z=0.84$ are significantly above the predicted ones with our model. The slope of the
\ha-halo mass relation increases drastically between 
$\lha \sim 10^{41.5}-10^{42} \lunits$ in the \citet{sobral10} data, and this is not reproduced in our model. 

It is important to notice that the observational samples of \citet{sobral10} might be affected by important systematics biasing their results. Firstly, 
their samples are corrected by dust attenuation by assuming that the extinction of the \ha\ line is simply $A_{H\alpha} = 1 {\rm mag}$. This is 
known to be an oversimplification \citep[see, e.g.][]{wild11}. The inferred spatial correlation function, used to derive the halo mass of the galaxy sample, relies critically in the redshift distribution of sources, which can also introduce a bias to the results. However, most importantly, the \citet{sobral10} samples might be affected by cosmic variance due to the small number of galaxies used to compute the correlation functions. \citet{orsi08} show that 
the effect of cosmic variance can be significantly larger than the statistical uncertainties that can be measured from the data. It is difficult to 
gauge the effect of cosmic variance in the samples of \citet{sobral10} without making a detailed analysis involving the construction of 
several mock catalogues of the observations. This is beyond the scope of the current work. Details on the procedure to construct
mock catalogues of narrow-band surveys can be found in \citet{orsi08}.

\section{Predictions for submillimetre observations}
\label{FIR}
\begin{figure}
\centering
\includegraphics[width=8.5cm]{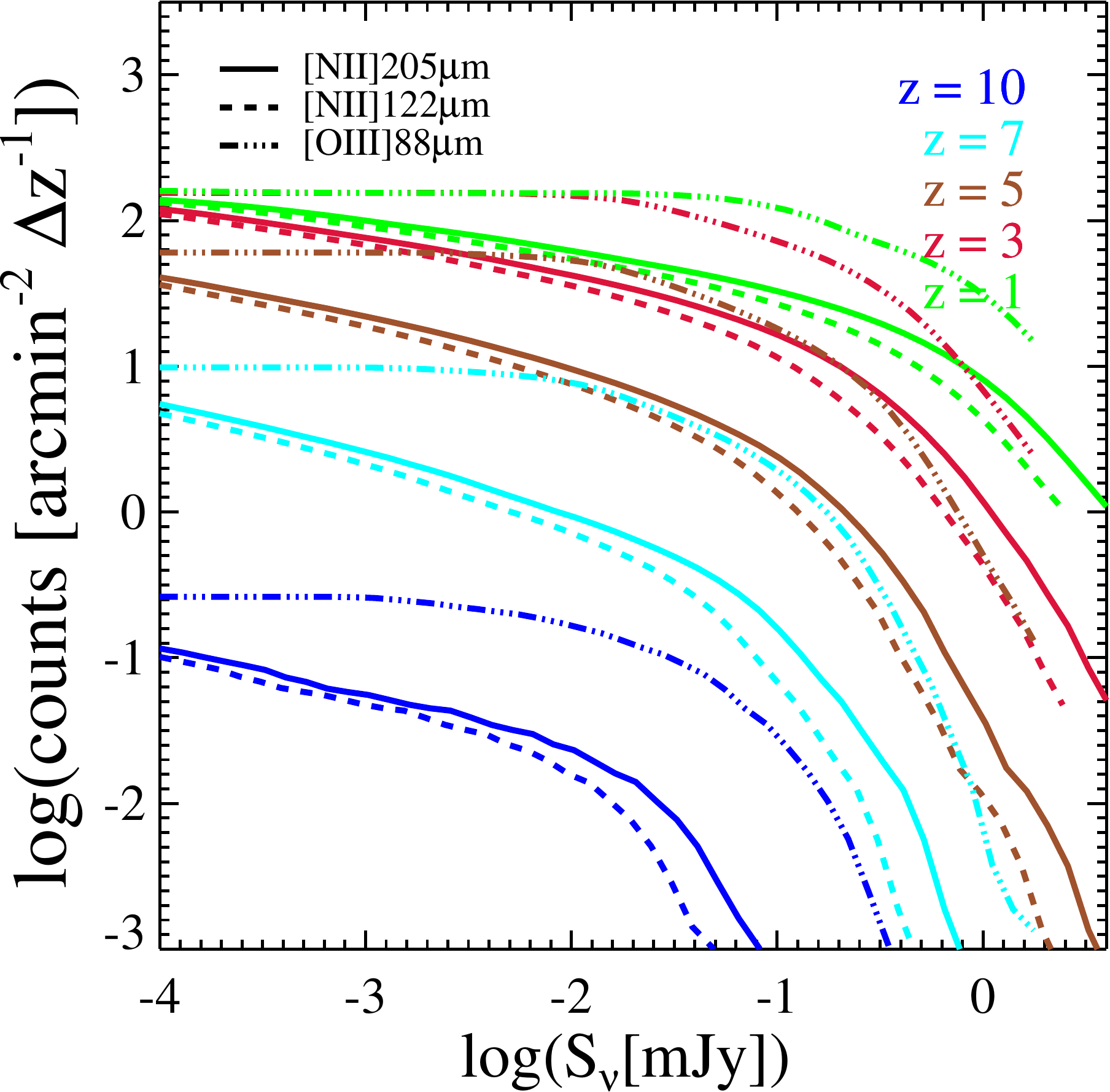}
\caption{The predicted number of \nii, \niiott\ and \oiiiee\ emitters normalised by redshift width as a function
of the limiting flux density for galaxies spanning the redshift range $1<z<10$. }
\label{fig.countsFIR}
\end{figure}

Searching for the first star forming galaxies is one of the key challenges for the new generation of millimetre and submillimetre facilities. 
The most appropriate tracers for these high redshift galaxies are FIR emission lines, such as \cii, 
\niiott, \nii\ and \oiiiee. These fine structure, collisional excitation lines provide 
cooling in regions of the ISM where allowed atomic transitions are not excited  \citep{decarli12}. Of all these FIR 
star-forming tracers the \cii\ line is the strongest one and it can account 
for about 0.1-1 per cent of the total FIR flux \citep{malhotra01}. Due to the low ionisation potential of carbon (11.3eV), \cii\ traces the 
neutral and ionised ISM. Since our model depends on the ionisation parameter of hydrogen (with an ionisation potential of 13.6eV), our 
predictions for the \cii\ do not account for the neutral regions where \cii\ is also produced. Thus, we cannot predict the complete
luminosity budget of \cii. Instead, we turn our attention to other FIR lines that can be targeted 
with submillimetre facilities. \nii\ and \niiott, for instance, can be accounted for in our model, since the ionisation potential of 
ionised nitrogen is 14.6eV, above hydrogen, and so these lines are produced in the ionised medium. 
The \nii\ is expected to become a recurrent target for high redshift objects since it has a nearly identical 
critical density for collisional excitation in ionised regions than \cii\ \citep{oberst06}. This makes the line ratio of \nii\ to \cii\ a 
sole function of the abundance between the two elements, regardless of the hardness of the ionising field. Therefore, 
both the star formation rate and the gas metallicity can be directly measured from observations of these two lines.

The \oiiiee\ line is also expected to be a recurrent target since it is significantly
stronger than the ionised nitrogen lines. 
Recently, by making use of a cosmological hydrodynamical simulation and a photo-ionisation model, \citet{inoue14} finds that 
the \oiiiee\ line is the best target to search for $z>8$ galaxies for submillimetre facilities such as ALMA.

We use our model to compute the number of high-redshift galaxies that could be detected by targeting these FIR lines.
Fig. \ref{fig.countsFIR} shows the number of \nii, \niiott\ and \oiiiee\ emitters predicted to be found in a square arc minute 
as a function of limiting flux density for different redshifts. An important caveat in this calculation is our assumption of
a constant density in the medium, $n_e = 10 {\rm [cm^{-3}]}$. The ratio $\nii/\niiott$, for instance, increases rapidly
towards higher values of the density of the gas when $n_e > 10 {\rm [cm^{-3}]}$ \citep{oberst06,decarli14}.

Furthermore, since our model does not compute the line profiles of these lines, in
order to obtain the density flux in units of ${\rm [mJy]}$ we assume for simplicity a top hat profile. The intrinsic 
line width is chosen to be $v = 50 {\rm [km/s]}$, consistent with the \nii\ width detected in \citep{oberst06}. 
Note that the y-axis is normalised by $\Delta z$, meaning
that the predicted number of objects at a given redshift will depend on the redshift range covered by the band
used to detect the line emission.

Overall, our model predicts that galaxies up to $z=10$ should be detectable in blank fields reaching depths that are
achievable by the current generation of submillimetre facilities. 
The beam size of ALMA, for instance, is of the order of a few arc seconds at most, so it would be 
necessary to perform of the order of hundreds of pointings and reach flux densities of $\sim 10^{-2} {\rm [mJy]}$ 
to detect very high redshift galaxies from a blank field.

Fig. \ref{fig.countsFIR} shows that \oiiiee\ is the strongest FIR line from the three shown, with surface number densities 
up to an order of magnitude larger than the two ${\rm [N II]}$ lines. At $z=10$, for instance, \oiiiee\ emitters could be detected
by surveying an area of the order of $\sim 10 {\rm [arcmin]^2}$ with a depth of $S_{\nu} \sim 10^{-2} {\rm [mJy]}$. Since 
\nii\ and \niiott\ emitters are predicted to be fainter than \oiiiee, the same flux and area is predicted to detect about the
same number of FIR ${\rm [N II]}$ emitters at $z=7$. At $z=10$, on the other hand, these would 
require a few hundred ${\rm [arcmin]^2}$ and a flux density of $\sim 100 {\rm [\mu Jy]}$ to be detected.

At $z=5$, FIR lines are predicted to be significantly easier to detect. 
A square arc minute survey with a flux depth of $\sim 0.01 {\rm [mJy]}$ 
should detect tens of \nii\ and \niiott\ emitters, and even a few hundreds of \oiiiee\ emitters per $\Delta z$. 
The number density of FIR line emitters grows by an order of magnitude by $z=3$ at the same flux depth, 
and about a factor 2 more by $z=1$.

Despite the simplifications made, Fig. \ref{fig.countsFIR} illustrates that a blind search of FIR lines in SF
galaxies down to flux depths and areas achievable with current instruments can
successfully result in a significant sample of high redshift objects to perform statistical analysis. Galaxies detected in this way
will define a population based only on the limiting flux of the line chosen to trace their SF activity. 

\section{Discussion}
\label{sec.discussion}
This paper presents the first model for emission line galaxies based on a fully-fledged 
semi-analytical model of galaxy formation in a hierarchical cosmology scenario. 
Our approach to model the different line luminosities is to invoke a simple, albeit meaningful, 
assumption that can lead to a physical interpretation of the predictions. 
Our model for the luminosities of emission line in star-forming galaxies relies on the critical assumption that
the ionisation parameter is directly related to the cold gas metallicity via Eq. \eqref{eq.qZ}. This assumption 
is supported by a number of observational and theoretical studies of line ratios of local star-forming galaxies
\citep[e.g.][and references therein]{nagao06,maier06,dopita06a,groves10,kewley13a}. However, 
it is likely that these two quantities are related to each other by a third, more fundamental physical property. 
For instance, the ionisation parameter depends on the gas pressure, which together with the hardness of the ionising 
radiation field determine the radii of the HII regions from which the ionisation, recombination and collisional
excitation processes take place. Hence, HII regions in a typical galaxy can display a distribution of radii, sizes and 
masses which, in turn, lead to a variety of local ionisation parameter values from which the line luminosities are 
produced \citep{dopita05,dopita06b,dopita06a}. 
The choice of a global ionisation parameter to represent the photoionisation process inside a galaxy could hence
be an oversimplification, meaning that a more sophisticated model might be required to account for internal structure
and dynamics of the ISM of each galaxy.

Despite the above, our simple approach has proven to predict the statistical properties
of emission lines, such as the BPT diagram (Fig. \ref{fig.bpt}) and the emission line LF evolution 
(Figs. \ref{fig.half},\ref{fig.oiiilf} and \ref{fig.oiilf}) in reasonable agreement with observations. 
Observational estimates also suggests that the ionisation parameter
of high redshift galaxies is higher with respect to low redshifts. Models and observations are also suggesting that
the line ratios observed in high redshift galaxies seem to imply ionisation parameters well above the upper 
limit of standard grids of photoionisation models,  $q \sim 10^{8}{\rm [cm/s]}$. 

Our model is constrained to lie in this range, so 
we cannot predict higher ionisation parameter values. However, it is encouraging that the model predicts increasingly
higher ionisation parameters for high redshifts, and that most of the galaxies in the modest SFR regime are predicted
to have the highest ionisation parameter the version of \mapp\ allows, i.e. $q = 4\times 10^{8} {\rm [cm/s]}$ (Fig. \ref{fig.QSFR}).

There has been recent evidence in the literature of an evolution of the BPT diagram with redshift \citep[e.g.][]{yeh13,kewley13b} 
from both an observational and theoretical stand point. This evolution is also consistent with the high values inferred
for the ionisation parameter of high redshift galaxies. In our model, the ionisation parameter is constrained to a set of values determined
by the gas metallicity, and so therefore there is no evolution in the shape of the BPT diagram at high redshifts, only a redistribution
of galaxies along the same sequence. 

Our model of galaxy formation also has limitations.
The most important one is related to the resolution limit of the dark matter haloes
resolved in the N-body simulation used. In this work, the minimum dark matter halo mass is $M_{\rm min} = 10^{10} {\rm [h^{-1}M_{\odot}]}$.
This limit impacts on semi-analytical models based on N-body merger trees \citep[e.g.][]{bower06}, creating
an artificial low abundance of faint galaxies
with low stellar masses. Only a fraction of these galaxies are hosted by haloes of the minimum halo mass, 
and the rest should be hosted by haloes with even lower halo mass, which are not resolved by the simulation. 
This artefact can be seen in the very faint end of the \ha\ luminosity functions, (Fig. \ref{fig.half}), and in the median halo mass
of faint line luminosities shown in Fig. \ref{fig.bias_mhalo}. Our results and conclusions are based on galaxies with \ha\ luminosities 
$L_{\rm H\alpha}>10^{40} \lunits$ and stellar masses $M_{\rm stellar} > 10^9 {\rm [h^{-1}M_{\odot}]}$, which 
selects galaxies above the halo mass resolution limit, and so should not be affected by this effect.

Another caveat of the model is the evolution of the predicted mass-metallicity relation. 
The predicted evolution in \sag\ of the metallicity of the stellar component, tightly related to the cold gas metallicity, is studied in
\citet{jimenez11} (see their Fig. 6). Overall, the stellar mass and metallicity of galaxies are lower at higher 
redshifts. However, unlike the apparent evolution of the mass-metallicity relation reported in observations of
high redshift galaxies \citep[e.g.][]{erb06,maiolino08,mannucci10} the \sag\ model shows no evolution in the 
normalisation of this relation. Galaxy evolution is predicted to increase the stellar component of galaxies and enrich their
ISM within the same sequence. This leads to a mass-metallicity relation 
for local galaxies that is already in place at high redshifts, in disagreement with what the observational estimates show.
Since the collisional excitation line strengths are sensitive to the gas-phase metallicity, this could explain
the partial disagreement of our model predictions with the observed \oiii\ LFs at high redshifts.

Despite the lack of evolution in the normalisation of the mass-metallicity relation, it is worth pointing out that 
the observationally inferred 
values for the gas metallicity, are subject to the biases and limitations imposed by the diagnostic used to infer the metallicities \citep{kewley08}. 
This is especially true when applying metallicity diagnostics to 
high redshift galaxies, where the ISM conditions are likely to be different from what is found in local galaxies \citep{cullen13}. 
The \sag\ model follows explicitly the chemical enrichment and 
evolution of 8 different 
chemical elements. Therefore,  the model computes an absolute value of the gas metallicity 
which is not straightforward to compare with the values inferred observationally. 

An extensive analysis of the mass-metallicity relation, its evolution and the role of the star-formation rate is beyond
of the scope of this paper and is discussed in Cora et al. (2014, in preparation).

\section{Conclusions}

\label{sec.conclusions}

Throughout this paper we explore the properties and evolution of emission line galaxies within 
a hierarchical cosmological model. This is the first attempt to model in detail the strong 
emission lines of galaxies in the optical and FIR range and to study their statistical properties
as a galaxy population. 

Our method consists in combining the semi-analytical model of galaxy formation \sag\ with the 
photoionisation and shock code \mapp\ to predict the emission line luminosities of galaxies. 
We make the critical assumption of relating the ionisation parameter of galaxies with the 
cold gas metallicity through Eq. \eqref{eq.qZ}. We choose the constant values in this relation 
to achieve a reasonable agreement of the BPT diagnostic and the \oii\ and \oiii\ LFs with the 
observational data available. We take advantage of the power of semi-analytical models
to explore the properties of the ELG population up to redshifts $z=10$, and present predictions 
of the detectability of FIR lines that can be targeted with millimetre and submillimetre facilities 
to explore the high redshift Universe.

We summarise our main findings as follows:

\begin{itemize}

\item Overall, our model predicts ELG luminosity functions in close agreement with current observational data. 
The model predicts an overabundance of \ha\ emitters at $z \lesssim 0.4$, but is reasonably consistent 
with the observed LFs in the range $0.9<z<2.2$. The luminosity functions of \oiii\ are find to be in partial agreement with observations. The abundance of \oii\ emitters at high redshifts is remarkably close to observed values, { but it is inconsistent with the observed abundance at $z=0$}. 
There are many reasons that can explain the disagreements between model predictions and observational measurements. 
First, our model relating the ionisation parameter $q$ with the cold gas metallicity $Z$ could be an oversimplification of the complex internal gas dynamics in HII regions of star-forming galaxies. Secondly, the ionisation parameter our model computes depends solely on the cold gas metallicity, which implies that if the chemical enrichment histories of galaxies are not realistic in our 
model then the resulting ionisation parameter values we obtain for high redshift galaxies would be incorrect. This would be reflected
in the luminosity function of \oii\ and \oiii. 
Finally, contamination from AGNs, shock excitation or LINERs might affect the observed luminosity functions 
in a rather complicated way. \citet{kewley13a} shows that the strong line ratios of star-forming galaxies and also AGNs are expected to evolve
with redshift, and that diagnostics such as the traditional BPT diagram cannot be applied to discriminate the source of emission lines in the
same way that this is done to local galaxies.

\item Our model predicts that star-forming galaxies at high redshifts have higher ionisation parameters than their 
counterparts at $z=0$. A similar 
conclusion has been suggested in a number of observations \citep[e.g.][]{brinchmann08,hainline09,nakajima13,richardson13}. This
is an expected result, since galaxies accumulate metals throughout their star-forming history, and so the bulk of the galaxy population 
at high redshifts is predicted to have less metal abundances than local galaxies. Since the ionisation parameter in our model scales 
inversely with the metallicity, high redshift galaxies have therefore higher values of the ionisation parameter.

\item We find that line luminosities can be used to infer the star-formation rate of galaxies, in spite of the scatter of the relation. 
We fit a polynomial functional form which depends on the emission line luminosity and the redshift of the galaxies. The scaling 
relations we present are in general accurate to within an order of magnitude for low luminosities and their accuracy improves
towards brighter line luminosities. The less favoured 
emission line to infer the SFR from is found to be \oiii.

\item We find that our model predicts modest values for the clustering of emission line galaxies, roughly consistent with observational estimates of \ha\ clustering. The bias of emission line galaxies is found to increase towards brighter galaxy samples, especially in high redshift emission line galaxies.\oiii\ emitters present a strong connection between the bias and the line luminosity, and this correlation is found to be weaker in \oii\ and \ha\ emitters.
Overall, ELGs are found to be hosted by dark matter haloes of mass 
$M_{\rm halo} \lesssim 10^{12} {\rm [h^{-1}M_{\odot}]}$. 

\item We explore the detectability of FIR emission lines with submillimetre observations. \oiiiee\ is found to be the strongest line, although 
we do not explore the more common target \cii. A dedicated survey with ALMA covering several square arc minutes is predicted to 
find a handful of \oiiiee\ emitters at $z=10$ and \nii\ emitters at $z=7$ if the flux density probes down to a depth of $0.01 {\rm [mJy]}$.  

\end{itemize}

This is our first paper of a series exploring the properties of strong optical and FIR nebular emission in galaxies. 
Several improvements can be made to the model to 
account for more physical processes. For instance, the mid-plane pressure of the ISM could be used to infer an effective ionisation parameter. 
A detailed study of the chemical enrichment of galaxies throughout cosmic 
time could also prove to be key in validating or ruling out our simple modelling of the ionisation parameter.

Regardless of the limitations, this work constitutes the first attempt to explore the properties of emission line galaxies in a 
cosmological setting. With the advent of ongoing and future large area surveys of emission line galaxies, 
a detailed understanding of this galaxy population will be regarded as crucial to extract information about the large scale structure
of the Universe, and the process of galaxy evolution.

\section*{Acknowledgements}

{ We acknowledge the comments from the anonymous referee, which helped improving the contents of this paper. We also acknowledge discussions with Chun Ly, Cedric Lacey, Carlton Baugh, Claudia Lagos, Iv\'an Lacerna, Paulina Troncoso and Alex Hagen.} 
We acknowledge the \sag\ collaboration for kindly allowing to use the code for the calculations presented here, and for the many
discussions that contributed to this paper.
AO gratefully acknowledges support from FONDECYT project 3120181. 
NP received support from Fondecyt Regular No. 1110328, BASAL PFB-06 ÓCentro de Astronomia y Tecnologias AfinesÓ. NP, TT and AO acknowledge support by the European Commissions Framework Programme 7, through the Marie Curie International Research Staff Exchange Scheme LACEGAL (PIRSES-GA-2010-269264). 
SC acknowledges funding from CONICET (PIP-220) and Agencia Nacional de Promoci\'on Cient\'ifica 
y Tecnol\'ogica (PICT-2008-0627), Argentina. TT acknowledges 
funding from GEMINI-CONICYT Fund Project 32090021, Comit\'e Mixto ESO-Chile and Basal-CATA (PFB$-06/2007$).
The calculations made in this paper were performed in the Geryon supercluster at Centro de Astro-Ingenier\'ia UC. 
The Anillo ACT-86, FONDEQUIP AIC-57, and QUIMAL 130008 provided funding for several improvements of the Geryon cluster.

\bibliographystyle{mn2e}
\bibliography{ref}

\end{document}